\documentclass{article}

\usepackage{PRIMEarxiv}

\usepackage[utf8]{inputenc} 
\usepackage[T1]{fontenc}    
\usepackage{hyperref}       
\usepackage{url}            
\usepackage{booktabs}       
\usepackage{amsfonts}       
\usepackage{nicefrac}       
\usepackage{microtype}      
\usepackage{lipsum}
\usepackage{fancyhdr}       
\usepackage{graphicx}       
\graphicspath{{media/}}     

\usepackage{tikz}
\usetikzlibrary{arrows,automata,positioning}

\usepackage{mathtools}
\usepackage{xspace}

\usepackage{hyperref}

\usepackage[utf8]{inputenc} 
\usepackage[T1]{fontenc}    
\usepackage{hyperref}       
\usepackage{url}            
\usepackage{booktabs}       
\usepackage{amsfonts}       
\usepackage{nicefrac}       
\usepackage{microtype}      
\usepackage{xcolor}         
\usepackage{answers}

\usepackage[ruled,vlined]{algorithm2e}

\title{Weighted Riesz Particles
}

\author{
  Xiongming Dai  \\
  Division of Computer Science and Engineering \\
  Louisiana State University \\
  Baton Rouge,LA70803, USA\\
  \texttt{\{xdai2\}@email} \\
   \And
  Gerald Baumgartner \\
  Division of Computer Science and Engineering \\
  Louisiana State University \\
  Baton Rouge,LA70803, USA\\
  \texttt{\{gb\}@email} \\
}

\begin{document}
\maketitle

\begin{abstract}

Markov chain Monte Carlo (MCMC) methods are simulated by local exploration of complex statistical distributions, and while bypassing the cumbersome requirement of a specific analytical expression for the target, this stochastic exploration of an uncertain parameter space comes at the expense of a large number of samples, and this computational complexity increases with parameter dimensionality. Although at the exploration level, some methods are proposed to accelerate the convergence of the algorithm, such as tempering, Hamiltonian Monte Carlo, Rao-redwellization, and scalable methods for better performance, it cannot avoid the stochastic nature of this exploration. 
We consider the target distribution as a mapping where the infinite-dimensional Eulerian space of the parameters consists of a number of deterministic submanifolds and propose a generalized energy metric, termed weighted Riesz energy, where a number of points is generated through pairwise interactions, to discretize rectifiable submanifolds.
We study the properties of the point, called Riesz particle, and embed it into sequential MCMC, and we find that there will be higher acceptance rates with fewer evaluations, we validate it through experimental comparative analysis from a linear Gaussian state-space model with synthetic data and a non-linear stochastic volatility model with real-world data.

\end{abstract}


\section{Introduction}

Markov chain Monte Carlo integration methods provide a feasible way for Bayesian analysis \cite{hastings1970monte,metropolis1953equation,geman1984stochastic,gelfand1990sampling}, which requires the evaluation of complex and often high-dimensional integrals to obtain posterior distributions for the unobserved latent quantities of interest in the model. The advantage of this method is that it guarantees convergence to the target distribution after a "burn-in" period \cite{robert1999monte}. This robustness may, however, lead to very low convergence rates in that the exploration of the uncertain space, meaning some part of the space supporting the target distribution that has a great probability mass under that distribution may take a long time, as the simulation usually traverses by local jumps in the vicinity of the current position \cite{gelman1997weak}.   

 Many discretization techniques in numerical analysis are based on finite samples that adequately cover the underlying space \cite{breger2018points}. Quasi-Monte Carlo design was studied for compact smooth Riemannian manifolds \cite{brandolini2010quadrature}. Points on a design space that derives from energy-based functions often have desirable separation, the worse-case error of integration bounds the covering radius and it provides asymptotically optimal covering radii\cite{damelin2005point,hardin2012quasi,hardin2005minimal,borodachov2008asymptotics,borodachov2014low}. This yields equilibrium points that are useful in a variety of applications, especially in high-dimensional sampling. We consider the target distribution as a mapping where the infinite-dimensional Eulerian space of parameters consists of several deterministic submanifolds. We propose a new flexible minimum energy metric to discretize rectifiable submanifolds via particle interaction. We study the characterization of deterministic points (called weighted Riesz particles) and  embed them into sequential
MCMC, and we find that there will be higher acceptance rates with fewer evaluations, we validate
our assertion through experimental comparative analysis from a linear Gaussian state-space model
with synthetic data and a non-linear stochastic volatility model with real-world data.



In this paper, we focus on analyzing our novel energy criterion and the characterization of particles generated based on this criterion, as well as on how to improve the acceptance rate of the MCMC with fewer evaluations. We present an efficient algorithm for deterministically sampling the target distribution with weighted Riesz energy minimization, from which particles sparsely represent integrable geometric manifolds, with a small number of samples to approximate the target posterior distribution. 


In Section 2, we briefly introduced the minimum energy model. Here, we propose a new generalized energy criterion and focus on asymptotic behavior, separation and coverage radius. In Section 3, we present a novel sampler with weighted Riesz particles, where the discretized deterministic submanifold inherits a special representation of the sampled space. We then describe the procedure for sequentially sampling weighted Riesz particles and embedding them in the Metropolis-Hastings algorithm for the hidden Markov model. In Section 4, we validate the algorithm through tracking the stochastic volatility and present its performance and error analysis. Section 5 summarizes our contributions.
\section{Weighted Riesz Energy Criterion}

In this section, we introduce the main idea of discrete minimum energies on rectifiable high-dimensional manifolds and propose a generalized energy criterion. Then, we study the asymptotic behavior of the corresponding configurations and representations of weighted Ritz particles in terms of separation and covering radii.

\subsection{Discrete Weighted Riesz Energy}
Let $\mathfrak{C}$ denote a compact set in $\mathbb{R}^d$ whose $d$-dimensional Borel measure, $\mathbb{B}_d(\mathfrak{C} )\subset (\mathfrak{C},\mathbb{R}^d)$, is finite, and $h$ denote a bi-Lipschitz mapping from $\mathfrak{C} \times \mathfrak{C} $ to $\mathbb{R}^d$, for a collection of $n(\geq 2)$ distinct points of configuration in $\mathfrak{C}$, let $X_{1:n}=\{x_1,...,x_n\}$, we define the energy of $X_{1:n}$ to be
\begin{equation}{\label{b2}}
E(X_{1:n}):=\sum_{i=1}^{n}\sum_{j=1,j\neq i}^{n}h(x_i,x_j)=\sum_{i\neq j}^{}h(x_i,x_j),
\end{equation}
and let 
\begin{equation}{\label{b3}}
\mathcal{E}(\mathfrak{C} ,n):=\text{inf}\{E(X_{1:n}):X_{1:n}\subset \mathfrak{C},\left | X_{1:n} \right |=n  \}
\end{equation}
be the minimal discrete $n$-point energy of the configuration in $\mathfrak{C}$, where $\left | X_{1:n} \right |$ represents the cardinality of the set $X_{1:n}$. (\uppercase\expandafter{\romannumeral1}) For $h(x_i,x_j)=-\text{log}\parallel x_i-x_j\parallel $, it was first proposed by M.Fekete who explored the connection between discretized manifolds and polynomial interpolation\cite{fekete1923verteilung}. In computational complexity theory, 
Smale \cite{smale1998mathematical} proposed the $7th$ problem in his list of "Mathematical problems for the next century", which is how to design a polynomial time algorithm for generating ``nearly" optimal logarithmic energy points $X_{1:n}^{*}$, also called Fekete points, on the unit sphere in $\mathbb{R}^3$ that satisfy $E(X_{1:n}^{*})-\mathcal{E}(\mathbb{S}^2,  n)\leq \footnote{\text{In this paper} { {$C_i$}} $\in \mathbb{R}^+,i=1,2,... $\text{denote different constants}.}{ {C_1}} \cdot \log n$ for some universal constant { {${\color{ red}C_1}$}} ; (\uppercase\expandafter{\romannumeral2}) when $h(x_i,x_j)=\frac{1}{\parallel x_i-x_j\parallel^s}, s\in \mathbb{R}^+$, let $\mathcal{E}_{s}(\mathfrak{C} ,n)$ denote the Riez $s$-energy, by Taylor's formula, for any $s\in (0,+\infty )$, we have
\begin{equation}{\label{boob}}
\begin{split}
\lim_{s\rightarrow 0^+}\mathcal{E} _{s}(\mathfrak{C} ,n)=\lim_{s\rightarrow 0^+}\frac{n(n-1)+s\mathcal{E} _{\text{log}}(\mathfrak{C} ,n)+\mathcal{O}(s)}{s}=\mathcal{E} _\text{log}(\mathfrak{C} ,n).
\end{split}
\end{equation}

Consequently, the Fekete points set $X_{1:n}^{(s)}$ can be considered as limiting cases of point sets that minimize the discrete Riez $s$-energy, which is widely used to discretize manifolds via particle interactions in Euclidean space \cite{borodachov2008asymptotics,hardin2004discretizing}.

 From the point of view of statistical high-dimensional sampling, we consider a sufficiently large $d$ and propose the minimally weighted Riesz energy criterion of

\begin{align}
\label{eqn:eqlabel10}
\begin{split}
 & \mathcal{E}_{\beta} (\mathfrak{C} ,n)=\min_{x_i,x_j}\left \{ \sum_{i=1}^{n-1}\sum_{j=i+1}^{n}\frac{\omega(x_i,x_j)}{\parallel x_i-x_j\parallel^s}  \right \}^{\frac{1}{s}}, 
 \omega (x_i,x_j)\propto e^{\left [ \alpha \cdot \kappa (x_i) \kappa (x_j) +\beta \cdot \parallel x_i-x_j\parallel \right ]^{-\frac{s}{2d}}}.\\
\end{split}
\end{align}
When $s\rightarrow \infty$, the formulation ~\eqref{eqn:eqlabel10} is convex, the denominator for $\mathcal{E}_{\beta} (\mathfrak{C},n)$ approximates $\parallel x_i-x_j\parallel$, thus, it inherits the properties of Riesz energy, termed as weighted Riesz energy criterion.
To obtain a finite collection of point sets distributed according to a specified non-uniform density, e.g. points that can be used as weighted integrals or for the design of complex surfaces, where more points are needed in regions of higher curvature, we introduce $\omega (x_i,x_j)$ in ~\eqref{eqn:eqlabel10}, where $\kappa (x)\propto -\ln{f(x)}$, $\parallel x_i-x_j\parallel$ is involved to ensure that is locally bounded for $\alpha=-1$, $\beta$ is a local discrepancy coefficient to balance off the local conflict with the distributed points when short-range interactions between points are the dominator. Here, $\beta>0$. Thus, given a proper distribution $f(x)$, we can use $\mathcal{E}_{\beta} (\mathfrak{C}, n)$ to generate a sequence of $n$-point configurations that are "well-separated" and have asymptotic distribution $f(x)$.
 
 Our weighted Riesz energy $\mathcal{E}_{\beta} (\mathfrak{C},n)$ is continuous and derivable with respect to the parameter $\beta \subset \mathbb{R}$ from ~\eqref{eqn:eqlabel10}, it provides a more flexible and versatile framework when we discretize the submanifolds via particle interactions.
\subsection{Asymptotics for Extremal Weighted Riesz Energy Criterion}
{\textbf {Properties of $\mathfrak{C}(x_i,x_j)$ .}} (\uppercase\expandafter{\romannumeral1}) $\omega(x_i,x_j)$ is continuous as a function of $\kappa (x)\propto -\ln{f(x)}$ when $\exists \beta_0>0$ satisfying $\beta \leq \beta_0$; it is a positive constant when $\exists \beta_1>0$ satisfying $\beta \geq \beta_1$; (\uppercase\expandafter{\romannumeral2}) There exists a neighborhood set $\mathfrak{C}'$, where $x_i',x_j'  \in \mathfrak{C}',  \omega(x_i',x_j')$ is bounded and larger than zero; (\uppercase\expandafter{\romannumeral3}) $\omega(x_i,x_j)$ is bounded on any closed and compact metric space $\mathfrak{C}$.

Assume the compact set $\mathfrak{C} \subset \mathbb{R}^d$, for high dimension $s>d$, we define the generalized Borel measure on sets $\mathbb{S} \subset \mathfrak{C}$ with $\mathcal{U}_d^s(\mathbb{S}) :=  \int _\mathbb{S} \omega(x_i,x_j)d\mathcal{U}_d(x).$
It is bounded and the corresponding normalized form: $u_d^s(\mathbb{S}):=  \mathcal{U}_d^s(\mathbb{S})/ \mathcal{U}_d^s({\mathfrak{C}}).
$

$\textbf{Measure Metric.}$ Consider a Euclidean space $\mathbb{R}^d, d \ge 2$, for $s>d$, let $\mu(\sigma \text{-algebra}):=  \cup_{d=2}^{\infty }\{ \parallel x_i-x_j\parallel^d\}$, represent a Borel measure from the $\sigma $-algebra on $\mathfrak{C}$, a measure $\phi  $ in $\mathfrak{C}_i$ is a non-negative $\sigma $-algebra set function defined on $\mu(\sigma \text{-algebra})$ and finite on all compact sets $\mathfrak{C}_i \subset \mathfrak{C},i\in [1,n]$. If $\phi  < \infty $, then the measure $\phi $ is called finite. Generally, for the smallest $\sigma $-algebra, containing all compact subsets of $\mathfrak{C}_i$. 

According to the measure theoretics \cite{hunter2011measure}, we have the following novel version of the Poppy-Seed Bagel Theorem \cite{borodachov2019discrete} for weighted Riesz energy.\\
{\textbf {Theorem 2.2.1. }} Given a distribution $f(x)$ with respect to $d$-rectifiable set $\mathfrak{C}$ embedded in Euclidean space, $\omega(x_i,x_j)>0$ is bounded and continuous on the closed Borel sets $\mathbb{S} \subset \mathfrak{C} \times \mathfrak{C} $, for $s>d$, the minimal weighted Riesz energy configuration on $\mathfrak{C}$ from $\mathcal{E}_{\beta}(\mathfrak{C},n)$ where the $n$-point interacts via the $h_{\beta}(x_i,x_j)$ potential, have
\begin{equation}\label{eqn:eqlabel13}
 \lim_{n\rightarrow \infty }\frac{\mathcal{E}_{\beta} (\mathfrak{C},n)}{n^{\frac{2}{s}+\frac{1}{d}}}=\frac{{\color{ red}C_2}}{[\mathcal{U}_d^s(\mathbb{S})]^{\frac{1}{d}}}.
 \end{equation}
 Moreover, if $\mathcal{U}_d^s(\mathbb{S})>0$, any configuration $X_{1:n},n>1$ generated by asymptotically minimizing weighted Riesz energy $\mathcal{E}_{\beta} (\mathfrak{C},n)$ is uniformly distributed with respect to $\mathcal{U}_d$, that is, 
\begin{equation}\label{eqn:eqlabel14}
\lim_{n\rightarrow \infty }\frac{1}{n}\sum_{i=1,i\neq j}^{n }\parallel x_i-x_j\parallel =u_d^s(\mathbb{S}).
\end{equation}
$\textbf{Proof of Theorem 2.2.1.}$ We divide the proof of Theorem 2.2.1 into two parts; The proof works via induction with Lemma 2.2.2 for ~\eqref{eqn:eqlabel13} {\color{red}
in the main text}, and Lemmas 2.2.3, 2.2.4, 2.2.5, 2.2.6 and 2.2.7 for ~\eqref{eqn:eqlabel14} {\color{red}
in the main text}.

$\textbf{Lemma 2.2.2.}$  Given a distribution $f(x)$ with respect to $d$-rectifiable set $\mathfrak{C}$ embedded in Euclidean space, $\omega(x_i,x_j)>0$ is bounded and continuous on the closed Borel sets $\mathbb{S} \subset \mathfrak{C} \times \mathfrak{C} $, for $s>d$ and $\beta \in(0,\beta_0]\cup [\beta_1,+\infty ),\beta_0,\beta_1,{\color{ red}C_3} \in \mathbb{R}$, the minimal weighted Riesz energy configuration on $\mathfrak{C}$ from $\mathcal{E}_{\beta}(\mathfrak{C},n)$ where the $n$-point interacts via the $h_{\beta}(x_i,x_j)$ potential, have
\begin{equation}\label{eqn:eqlabel13dd}
 \lim_{\substack{\mathllap{n} \to \infty \\ \mathllap{\beta^{-}} \to \beta_0}}\frac{\mathcal{E}_{\beta} (\mathfrak{C} ,n)}{n^{\frac{2}{s}+\frac{1}{d}}}=\frac{{ {{\color{ red}C_3}}}}{[\mathcal{U}_d^s(\mathbb{S})]^{\frac{1}{d}}},\ \ \lim_{\substack{\mathllap{n} \to \infty \\ \mathllap{\beta^{+}} \to \beta_1}}\frac{\mathcal{E}_{\beta} (\mathfrak{C} ,n)}{n^{\frac{2}{s}+\frac{1}{d}}}=\frac{{ {{\color{ red}C_3}}}}{[\mathcal{U}_d^s(\mathbb{S})]^{\frac{1}{d}}}. 
\end{equation}
$\textbf{Proof}$ $\mathcal{E}_{\beta}(\mathfrak{C}, n)$ is strictly decreasing as $\beta$ increases, this monotonicity makes it possible to analyze the asymptotics and extend it into high-dimensional sampling on the compact set $\mathfrak{C}$ under mild assumptions. Let $J(\beta'):=\lim_{n\rightarrow \infty }\lim_{\beta \rightarrow \beta' }\mathcal{E}_{\beta} (\mathfrak{C},n)$, 
 \begin{equation*}{\label{h12}}
J(\beta):=\sum_{i=1}^{n}\sum_{j=1,j\neq i}^{n}\left[h_{\beta}(x_i,x_j)\right ]^{\frac{1}{s}}.
\end{equation*}
$h_{\beta}(x_i,x_j)$ is also strictly decreasing as $\beta$ increases, we firstly focus on $\beta \in(0 ,\beta_0]\cup [\beta_1,+\infty ),\beta_0,\beta_1 \in \mathbb{R}^+$ , then relax this assumption later, define
\begin{equation*}\label{eqn:eqlabel18}
h_{\beta'}(x_i,x_j):=\lim_{\beta \rightarrow \beta' }\frac{ \omega(x_i,x_j)}{\parallel x_i-x_j\parallel^s},\omega(x_i,x_j)> 0,
\end{equation*}
if $\beta \leq \beta_0$ is sufficiently small such that
\begin{equation}\label{eqn:eqlabel19a}
\kappa (x_i)\kappa (x_j)\gg\beta_0 \cdot \parallel x_i-x_j\parallel,
\end{equation}
then,
\begin{equation}\label{eqn:eqlabel19}
h_{\beta_0^{-}}(x_i,x_j):=\lim_{\beta\rightarrow \beta_0^{-}}h_{\beta}(x_i,x_j)=\frac{e^{\left [- \kappa (x_i) \kappa (x_j) \right ]^{-\frac{s}{2d}}}}{\parallel x_i-x_j\parallel^s}.
\end{equation}
From Taylor's theorem
\begin{equation*}\label{eqn:eqlabel19v}
e^z=1+z+\frac{z^2}{2!}+\cdots +\frac{z^{k'}}{k'!},k'\to \infty, z\in \mathbb{R}.
\end{equation*}
Let $z={\left [- \kappa (x_i) \kappa (x_j) \right ]^{-\frac{s}{2d}}}$, substitute ~\eqref{eqn:eqlabel19a} into ~\eqref{eqn:eqlabel19}, 
\begin{equation*}
\label{eqn:eqlabel20d2a}
\begin{split}
h_{\beta_0^{-}}(x_i,x_j)&=\frac{1+z+\frac{z^2}{2!}+\cdots +\frac{z^{k'}}{k'!}}{\parallel x_i-x_j\parallel^s}\ge \frac{1}{\parallel x_i-x_j\parallel^s}+ \frac{\left[-\beta_0 \cdot \parallel x_i-x_j\parallel\right]^{-\frac{s}{2d}}}{\parallel x_i-x_j\parallel^s}+\\
&...+\frac{\left[-\beta_0 \cdot \parallel x_i-x_j\parallel\right]^{\frac{-{s{k'}}}{2d}}}{{k'}!\parallel x_i-x_j\parallel^s}.
\end{split}
\end{equation*}
For $s>d$, the right-hand side terms belong to the classical Riesz-kernel model, from  the Poppy-Seed Bagel Theorem \cite{borodachov2019discrete}, there exists a ${{\color{ red}C_4}}$,
\begin{equation*}\label{eqn:eqlabeld19v}
h_{\beta_0^{-}}(x_i,x_j)=\frac{{\color{ red}C_4}}{[\mathcal{U}_d^s(\mathbb{S})]^{\frac{s}{d}}} \cdot n^{1+\frac{s}{d}}\cdot n.
\end{equation*}
Thus,
\begin{equation*}\label{eqn:eqlabeld19dddv}
J(\beta_0^{-}):=\sum_{i=1}^{n}\sum_{j=1,j\neq i}^{n}\left[h_{\beta_0^{-}}(x_i,x_j)\right ]^{\frac{1}{s}}=\frac{{ {C_8}}}{[\mathcal{U}_d^s(\mathbb{S})]^{\frac{1}{d}}}\cdot n^{\frac{1}{s}+\frac{1}{d}}\cdot n^{\frac{1}{s}}.
\end{equation*}
Similarly, if $\beta \geq \beta_1$ is  sufficiently large such that $ \kappa (x_i) \kappa (x_j) \ll \beta_1 \cdot \parallel x_i-x_j\parallel$, then
\begin{align}
\label{eqn:eqlabel20ew}
\begin{split}
   h_{\beta_1^{+}}(x_i,x_j) :=&\lim_{\beta\rightarrow \beta_1^{+}}h_{\beta}(x_i,x_j)=\frac{e^{\left [ \beta_1 \parallel x_i-x_j\parallel \right ]^{-\frac{s}{2d}}}}{\parallel x_i-x_j\parallel^s}
   = \frac{1}{\parallel x_i-x_j\parallel^s}+\\
   &\frac{\left[\beta_1 \cdot \parallel x_i-x_j\parallel\right]^{-\frac{s}{2d}}}{\parallel x_i-x_j\parallel^s}+... + \frac{\left[\beta_1 \cdot \parallel x_i-x_j\parallel\right]^{\frac{-{s{k'}}}{2d}}}{{k'}!\parallel x_i-x_j\parallel^s}.
\end{split}
\end{align}
It provides a flexible framework to prove the asymptotics of the proposed weighted Riesz energy criterion for ~\eqref{eqn:eqlabel20ew} that we will frequently refer to it for the following lemma and related proof.

For $s>d$, the right-hand side terms belong to the classical Riesz-kernel model, from  the Poppy-Seed Bagel Theorem \cite{borodachov2019discrete}, there exists a ${{\color{ red}C_5}}$,
\begin{equation*}\label{eqn:eqlabeld1dd9v}
h_{\beta_1^{+}}(x_i,x_j)=\frac{{\color{ red}C_5}}{[\mathcal{U}_d^s(\mathbb{S})]^{\frac{s}{d}}} \cdot n^{1+\frac{s}{d}}\cdot n.
\end{equation*}
Thus,
\begin{equation*}\label{eqn:eqlabeld19dvd}
J(\beta_1^{+}):=\sum_{i=1}^{n}\sum_{j=1,j\neq i}^{n}\left[h_{\beta_1^{+}}(x_i,x_j)\right ]^{\frac{1}{s}}=\frac{{ {C_7}}}{[\mathcal{U}_d^s(\mathbb{S})]^{\frac{1}{d}}}\cdot n^{\frac{1}{s}+\frac{1}{d}}\cdot n^{\frac{1}{s}}.
\end{equation*}
As $J({\beta})$ is strictly decreasing, and continuous and derivative for $\beta\in \mathbb{R}^+$, Consequently, There exists a ${{\color{ red}C_3}}$, 
\begin{equation*}\label{eqn:eqlabel1dx3}
 \lim_{n\rightarrow \infty }\frac{\mathcal{E}_{\beta} (\mathfrak{C} ,n)}{n^{\frac{2}{s}+\frac{1}{d}}}=\frac{{\color{ red}C_3}}{[\mathcal{U}_d^s(\mathbb{S})]^{\frac{1}{d}}}.
 \end{equation*}
Thus, ~\eqref{eqn:eqlabel13dd} holds.

From Lemma 2.2.2, as $n \to \infty$, the approximation of $\mathcal{E}_{\beta} (\mathfrak{C} ,n)$ is not correlated with $\beta$.
That is, we are assuming that $\beta$ approximates a specific real value, and for the convenience of introducing Taylor's theorem to derive, it does not affect the final limit value of $\mathcal{E}_{\beta} (\mathfrak{C},n)$ for $n \to \infty$.
  
Analogous to the proof of classical Poppy-Seed Bagel Theorem \cite{borodachov2019discrete}, we define 
\begin{equation*}\label{eqn:eqlabel17a}
\mathcal{T}(d):= 2+\frac{s}{d},n \geq 2.
\end{equation*}
$\lambda (n):=n^{\mathcal{T}(d)}$, for $n\geq 2$, $\lambda (1):=1$.
And define
\begin{equation}\label{eqn:eqlabel17}
\psi_{s,d} (\mathfrak{C}):=\lim_{n\rightarrow \infty }\frac{\mathcal{E}^s_\beta (\mathfrak{C} ,n)}{\lambda (n)},
\end{equation}
let $\psi_{s,d}^{\text{inf}} (\mathfrak{C})={\text{inf}}(\psi_{s,d} (\mathfrak{C}))$, $\psi_{s,d}^{\text{sup}} (\mathfrak{C})={\text{sup}}(\psi_{s,d} (\mathfrak{C}))$ and decompose the $d$-rectifiable set $\mathfrak{C}$ into different subsets $\mathfrak{C}_i, i \in \mathbb{R}^+$, satisfying $\cup_{i=1}^\infty {\mathfrak{C}_i}=\mathfrak{C}$. 
\\
{\textbf{Lemma 2.2.3.}} \cite{borodachov2019discrete} $\exists \alpha_1, \alpha_2 \in \mathbb{R}^+$,  $\mathcal{T}(d)$ is continuous and derivative for $d \in \mathbb{R}^+$, the function $U(t)=\alpha_1t^{\mathcal{T}(d)-1}+\alpha_2(1-t)^{\mathcal{T}(d)-1}$
has the minimum for $t \in [0,1]$ where occurs at the points $t^*:=\frac{1}{1+(\frac{\alpha_1}{\alpha_2})^\frac{1}{\mathcal{T}(d)-2}}$ with $U(t^*)= \left[ \alpha_2^{\frac{-1}{\mathcal{T}(d)-2}}+\alpha_1^{\frac{-1}{\mathcal{T}(d)-2}}\right]^{2-\mathcal{T}(d)}$.
\\
The proof is straightforward from the first order derivative of the function $\frac{dU(t)}{dt}=0$.

We will introduce the subadditivity and superadditivity properties as follows.

{\textbf{Lemma 2.2.4.}} $\exists \mathfrak{C}_j,\mathfrak{C}_k \subset \mathfrak{C}$, and $\mathfrak{C}_j,\mathfrak{C}_k \not\subset  \emptyset $, $j\neq k$, $\mathcal{T}(d)$ is continuous and derivative for  $d \in \mathbb{R}^+$, let ${X_{1:n}^{(i)}},i,j$ and $k\in \mathbb{R}^+$ be an infinite sequence of $n$-point configurations on $\mathfrak{C}_j \cup \mathfrak{C}_k$ and define the unit $x_{1:n}^{(i)}$ that belongs to the compact and closed subset $\mathfrak{C}_j$ with probability 
\begin{equation*}\label{eqn:eqlabel2afgt}
{p_j} :=P(x_{1:n}^{(i)} \in \mathfrak{C}_j).
\end{equation*}
Assume that both $\psi_{s,d}^{{\text{inf}}} (\mathfrak{C}_j)$ and $\psi_{s,d}^{{\text{inf}}} (\mathfrak{C}_k)$ are bounded, for $s >d$, we have
\begin{equation}\label{eqn:eqlabel2b}
\lim_{n\rightarrow \infty }{\text{inf}}\left [ \frac{\mathcal{E}^s_\beta ({X_{1:n}})}{n^{\mathcal{T}(d)}} \right ]
\geq U(p_j),
\end{equation}
where $U(p_j):=\psi_{s,d}^{{\text{inf}}} (\mathfrak{C}_j)p_j^{\mathcal{T}(d)-1}+\psi_{s,d}^{{\text{inf}}} (\mathfrak{C}_k)(1-p_j)^{\mathcal{T}(d)-1}$.\\
$\textbf{Proof}$ we follow an argument in \cite{borodachov2019discrete} and define $N_j:=\left \lfloor p_j\cdot n \right \rfloor$ the units that belong to $\mathfrak{C}_j$, for $\mathfrak{C}_k$, $N_k:=\left \lfloor(1-p_j)\cdot n \right \rfloor$. Here, we assume $n\rightarrow \infty $,
\begin{equation*}
\label{eqn:eqlabel2atku}
\begin{split}
  E^s_{\beta} ({X_{1:n}} ,n) & \geq E^s_{\beta} ({X_{1:n}^{(i)}} \cap \mathfrak{C}_j,N_j)+E^s_{\beta} ({X_{1:n}^{(i)}} \setminus{\mathfrak{C}_j},N_k) \geq \mathcal{E}^s_{\beta}(\mathfrak{C}_j,N_j)+\mathcal{E}^s_{\beta}(\mathfrak{C}_k,N_k).
\end{split}
\end{equation*}
We introduce the decomposed units into $\mathfrak{C}_j$ and $\mathfrak{C}_k$, and combine ~\eqref{eqn:eqlabel17}, let $\psi_{s,d} ({X_{1:n}}):=\mathcal{E}^s_\beta ({X_{1:n}},n) /n^{\mathcal{T}(d)}$, then
\begin{equation*}
\label{eqn:eqlabel20fa}
\begin{split}
   \lim_{n\rightarrow \infty } {\text{inf}} \left [ \psi_{s,d} ({X_{1:n}})\right ]&\geq\lim_{n\rightarrow \infty }{\text{inf}}\left [ \psi_{s,d} (\mathfrak{C}_j)\cdot(\frac{N_j}{n})^{\mathcal{T}(d)} \right ] +\lim_{n\rightarrow \infty }{\text{inf}}\left [ \psi_{s,d} (\mathfrak{C}_k)\cdot(\frac{N_k}{n})^{\mathcal{T}(d)} \right ]\\
  & \geq \psi_{s,d}^{{\text{inf}}} (\mathfrak{C}_j)p_j^{\mathcal{T}(d)-1}+\psi_{s,d}^{{\text{inf}}} (\mathfrak{C}_k)(1-p_j)^{\mathcal{T}(d)-1}=U(p_j).
\end{split}
\end{equation*}
If $p_j=0$, we have $N_k\rightarrow n$. Since
\begin{equation*}\label{eqn:eqlabel32b}
E^s_{\beta} ({X_{1:n}} ,n)\geq E^s_{\beta} ({X_{1:n}}\setminus \mathfrak{C}_j)\geq \mathcal{E}^s_{\beta} (\mathfrak{C}_k,N_k),
\end{equation*}
motivated by Lemma 2.2.3, we have
\begin{equation*}
\label{eqn:eqlabel202a}
\begin{split}
\lim_{n\rightarrow \infty }{\text{inf}}\left [ \frac{E^s_{\beta}(X_{1:n}) }{n^{\mathcal{T}(d)}}\right ]& \geq \lim_{n\rightarrow \infty }{\text{inf}}\left [ \frac{\mathcal{E}^s_{\beta}(\mathfrak{C}_k,N_k) }{n^{\mathcal{T}(d)}}\cdot (\frac{N_k}{n})^{\mathcal{T}(d)}\right ]\geq \psi_{s,d}^{{\text{inf}}} (\mathfrak{C}_k)=U(0).
\end{split}
\end{equation*}
Similarly, for $p_j=1$,
\begin{equation*}\label{eqn:eqlabel32ab}
\lim_{n\rightarrow \infty }{\text{inf}}\left [ \frac{E^s_{\beta}(X_{1:n}) }{n^{\mathcal{T}(d)}}\right ]\geq U(1).
\end{equation*}
Thus, ~\eqref{eqn:eqlabel2b} holds. \\
{\textbf{Lemma 2.2.5.}} $\exists \mathfrak{C}_j,\mathfrak{C}_k \subset \mathfrak{C}$, and $\mathfrak{C}_j,\mathfrak{C}_k \not\subset  \emptyset $, $j\neq k$, $\mathcal{T}(d) > 2$ is continuous and derivative for  $d \in \mathbb{R}^+$, let $\alpha_3=\frac{1}{2-\mathcal{T}(d)}$, for $s >d$,
\begin{equation}\label{eqn:eqlabel27}
\psi_{s,d}^{\text{sup}} (\mathfrak{C}_j \cup \mathfrak{C}_k)^{\alpha_3} \ge \psi_{s,d}^{\text{sup}} (\mathfrak{C}_j)^{\alpha_3}+\psi_{s,d}^{\text{sup}} (\mathfrak{C}_k)^{\alpha_3}.
\end{equation}
{\textbf{Proof}}
If $\psi_{s,d}^{\text{sup}}(\mathfrak{C}_j)$ or $\psi_{s,d}^{\text{sup}} (\mathfrak{C}_k)$ equals zero, or one of the quantities $\psi_{s,d}^{\text{sup}}(\mathfrak{C}_j)$ or $\psi_{s,d}^{\text{sup}} (\mathfrak{C}_k)$ approximates infinite, as the size of set increase, $\mathcal{E}_{\beta}^s (\mathfrak{C} ,n)$ will increase, the lemma holds.

Hereafter, we follow an argument in \cite{borodachov2019discrete} and consider the general case of $\psi_{s,d}^{\text{sup}} (\mathfrak{C}_j)\in (0,\infty ),\psi_{s,d}^{\text{sup}} (\mathfrak{C}_k)\in (0,\infty )$, the distance of two set is defined with $r:=\left \| a_i-b_j \right \|,a_i \in \mathfrak{C}_j,b_j \in \mathfrak{C}_k,i,j\in \mathbb{R}^+$. Motivated by Lemma 2.2.4 with $\alpha_1=\psi_{s,d}^{\text{sup}}(\mathfrak{C}_j)$ and $\alpha_2=\psi_{s,d}^{\text{sup}} (\mathfrak{C}_k)$, for a given $n$ units, ${X_{1:n}^{(i)}} \cap \mathfrak{C}_j$ and ${X_{1:n}^{(i)}} \setminus{\mathfrak{C}_j}$ be configurations of $N_j:=\left \lfloor \tilde{p}\cdot n \right \rfloor$ and $N_k:=n-N_j$ points respectively such that $E^s_{\beta} ({X_{1:n}^{\mathfrak{C}_j}})< \mathcal{E}^s_{\beta} (\mathfrak{C}_j,N_j)+{ {{\color{ red}C_6}}}$ and $E^s_{\beta} ({X_{1:n}^{\mathfrak{C}_k}})< \mathcal{E}^s_{\beta} (\mathfrak{C}_k,N_k)+{ {{\color{ red}C_6}, {\color{ red}C_6}}} \in \mathbb{R}^+$, where
\begin{equation*}\label{eqn:eqlabel2adfg}
\tilde{p}=\frac{\psi_{s,d}^{\text{sup}} (\mathfrak{C}_j)^{\alpha_3}}{\psi_{s,d}^{\text{sup}} (\mathfrak{C}_j)^{\alpha_3}+\psi_{s,d}^{\text{sup}} (\mathfrak{C}_k)^{\alpha_3}}.
\end{equation*}
Then
\begin{equation*}
\label{eqn:eqlabel2a2a}
\begin{split}
\mathcal{E}^s_{\beta} (\mathfrak{C}_j\cup \mathfrak{C}_k ,n) &\leq E^s_{\beta}(X_{1:n}^{\mathfrak{C}_j}\cup X_{1:n}^{\mathfrak{C}_k}) =E^s_{\beta}(X_{1:n}^{\mathfrak{C}_j})+E^s_{\beta}( X_{1:n}^{\mathfrak{C}_k})+2\sum_{\substack{x\in X_{1:n}^{\mathfrak{C}_j} \\ y\in X_{1:n}^{\mathfrak{C}_k}}}\frac{ \omega(x,y)}{\left \| x-y \right \|}\\
& \leq \mathcal{E}^s_{\beta} (\mathfrak{C}_j,N_j)+ \mathcal{E}^s_{\beta} (\mathfrak{C}_k,N_k)+{ {{\color{ red}C_7}}}+2\cdot \frac{n^2}{r}\cdot \left \| \omega \right \|_{\mathfrak{C}_j\times\mathfrak{C}_k},
\end{split}
\end{equation*}
where $\left \| \omega \right \|_{\mathfrak{C}_j\times\mathfrak{C}_k}$ denotes the supremum of $\mathfrak{C}$ over $\mathfrak{C}_j \times \mathfrak{C}_k$. Dividing by $\lambda (n)$ and taking into account that $\lambda(N_j)/\lambda(n)\rightarrow {(\tilde{p})^{\mathcal{T}(d)}}$ as $n \rightarrow \infty$, we obtain
\begin{align*}
\label{eqn:eqlabel20a}
\begin{split}
   \psi_{s,d}^{\text{sup}} (\mathfrak{C}_j \cup \mathfrak{C}_k)  &=\lim_{n\rightarrow \infty }{\text{sup}}\frac{\mathcal{E}^s_{\beta} (\mathfrak{C}_j\cup \mathfrak{C}_j ,n)}{n^{\mathcal{T}(d)}} \le \lim_{n\rightarrow \infty }{\text{sup}}\frac{\mathcal{E}^s_{\beta} (\mathfrak{C}_j,N_j)}{n^{\mathcal{T}(d)}}+ \lim_{n\rightarrow \infty }{\text{sup}}\frac{\mathcal{E}^s_{\beta} (\mathfrak{C}_k,N_k)}{n^{\mathcal{T}(d)}} \\
   &= \lim_{n\rightarrow \infty }{\text{sup}}\frac{\mathcal{E}^s (\mathfrak{C}_j,N_j)}{N_j^{\mathcal{T}(d)}}{(\tilde{p})^{\mathcal{T}(d)}}+ \lim_{n\rightarrow \infty }{\text{sup}}\frac{\mathcal{E}^s (\mathfrak{C}_k,N_k)}{N_k^{\mathcal{T}(d)}}{(1-\tilde{p})^{\mathcal{T}(d)}} \\
    &\le \psi_{s,d}^{\text{sup}} (\mathfrak{C}_j)\cdot (\tilde{p})^{\mathcal{T}(d)}+\psi_{s,d}^{\text{sup}} (\mathfrak{C}_k)\cdot (1-\tilde{p})^{\mathcal{T}(d)}\\
    &=\left [ \psi_{s,d}^{\text{sup}} (\mathfrak{C}_j)^{\alpha_3}+\psi_{s,d}^{\text{sup}} (\mathfrak{C}_k)^{\alpha_3} \right ]^{\frac{1}{\alpha_3}}.
\end{split}
\end{align*}
Thus, Lemma 2.2.5 holds.\\
{\textbf{Lemma 2.2.6.}} $\exists \mathfrak{C}_j,\mathfrak{C}_k \subset \mathfrak{C}$, and $\mathfrak{C}_j,\mathfrak{C}_k \not\subset  \emptyset $, $j\neq k$, $\mathcal{T}(d) > 2$ is continuous and derivative for  $d \in \mathbb{R}^+$, let $\alpha_3=\frac{1}{2-\mathcal{T}(d)}$ for $s >d$,
\begin{equation}\label{eqn:eqlabel27dd}
\psi_{s,d}^{\text{inf}} (\mathfrak{C}_j \cup \mathfrak{C}_k)^{\alpha_3} \le \psi_{s,d}^{\text{inf}} (\mathfrak{C}_j)^{\alpha_3}+\psi_{s,d}^{\text{inf}} (\mathfrak{C}_k)^{\alpha_3}
\end{equation}
Furthermore, if $\psi_{s,d}^{{\text{inf}}} (\mathfrak{C}_j),\psi_{s,d}^{{\text{inf}}} (\mathfrak{C}_k) \ge 0$ and at least one of these oracles is finite, then for any infinite subset $N'$ of $n$ and any sequence $\{X_{1:n}\}_{n\in N'}$ of $n$-point configurations in $\mathfrak{C}_j\cup \mathfrak{C}_k$ such that
\begin{equation}\label{eqn:eqlabelx7}
\lim_{n\rightarrow \infty }\frac{\mathcal{E}^s_{\beta} (X_{1:n} ,n)}{\lambda (n)}=(\psi_{s,d}^{\text{inf}} (\mathfrak{C}_j)^{\alpha_3}+\psi_{s,d}^{\text{inf}} (\mathfrak{C}_k)^{\alpha_3})^\frac{1}{\alpha_3}
\end{equation}
holds, we have
\begin{equation}\label{eqn:eqlabel2adfjn}
p_j=\frac{\psi_{s,d}^{\text{inf}} (\mathfrak{C}_j)^{\alpha_3}}{\psi_{s,d}^{\text{inf}} (\mathfrak{C}_j)^{\alpha_3}+\psi_{s,d}^{\text{inf}} (\mathfrak{C}_k)^{\alpha_3}}.
\end{equation}
$\textbf{Proof}$ We assume $\psi_{s,d}^{\text{inf}} (\mathfrak{C})$ is bounded on the compact set $\mathfrak{C}_j$ and $\mathfrak{C}_k$. Let $\hat{N}$ be an infinite subsequence such that
\begin{equation*}\label{eqn:eqlabel24a}
\lim_{n\rightarrow \hat{N} }\frac{\mathcal{E}^s_{\beta} (\mathfrak{C}_j \cup \mathfrak{C}_k ,n)}{n^{\mathcal{T}(d)}}:=\psi_{s,d}^{\text{inf}} (\mathfrak{C}_j \cup \mathfrak{C}_k).
\end{equation*}
If $X_{1:n}^{(i)}$ is a sequence of $n$-point configurations on $\mathfrak{C}_j \cup \mathfrak{C}_k$ such that $E^s_{\beta}(X_{1:n}) \leq \mathcal{E}^s_{\beta} (\mathfrak{C}_j \cup \mathfrak{C}_k ,n)+{ {C_{11},C_{11}}} \in \mathbb{R}^+$. Similar to the proof of Lemma 2.2.4, if at least one of the oracles $\psi_{s,d}^{\text{inf}} (\mathfrak{C}_j)$ or $\psi_{s,d}^{\text{inf}} (\mathfrak{C}_k)$ is infinite, where $p_j=0$ or $1$. Generally, for $p_j \in (0,1)$, by Lemma 2.2.4, we obtain
\begin{equation*}\label{eqn:eqlabel246a}
\psi_{s,d}^{{\text{inf}}} (\mathfrak{C}_j \cup \mathfrak{C}_k)=\lim_{n\rightarrow \hat{N} } \frac{E^s_{\beta}(X_{1:n}) }{n^{\mathcal{T}(d)}}\geq U(p_j).
\end{equation*}
We follow an argument in \cite{borodachov2019discrete} and let 
\begin{equation*}\label{eqn:eqlabela2a}
\alpha'=\frac{\psi_{s,d}^{\text{inf}} (A)^{\alpha_3}}{\psi_{s,d}^{\text{inf}} (A)^{\alpha_3}+\psi_{s,d}^{\text{inf}} (B)^{\alpha_3}}
\end{equation*}
and $\{X_{1:n}\}_{n\in N'}$ be any sequence of $n$-point sets in $A\cup B$ satisfying
~\eqref{eqn:eqlabelx7}. Let $n \subset \hat{N}$ be any infinite subsequence on $\mathfrak{C}_j \cup \mathfrak{C}_k$ such that the quantity $p_j\geq 0$ as $n \rightarrow  \infty$, $n \in \hat{N}$. From Lemma 2.2.5 and ~\eqref{eqn:eqlabelx7}, we have
\begin{equation*}\label{eqn:eqlabela2afy1}
U(\alpha')=\lim_{n\rightarrow \infty }\frac{\mathcal{E}^s_{\beta} (X_{1:n} ,n)}{\lambda (n)}\geq U(p_j),
\end{equation*}
where $U(\alpha')$ is equal to the right-hand side of ~\eqref{eqn:eqlabelx7}, obviously, $\alpha'$ is the only minimum point of $U(\cdot)$. Consequently, $p_j=\alpha'$, ~\eqref{eqn:eqlabel2adfjn} holds.
If $X_{1:n}$ is a subsequence on $N':=\mathfrak{C}_j \cup \mathfrak{C}_k\setminus \hat{N}$, we denote the corresponding units that belong to the compact subset with probability $p_j'$, then
\begin{align*}
\label{eqn:eqlabel2aa}
\begin{split}
  \psi_{s,d}^{{\text{inf}}} (\mathfrak{C}_j \cup \mathfrak{C}_k) & =\lim_{n\rightarrow N' } \frac{E^s_{\beta}(X_{1:n}) }{n^{\mathcal{T}(d)}}\geq U(p'_j)
   \geq U(\alpha')\\
   &=\left [\psi_{s,d}^{\text{inf}} (\mathfrak{C}_j)^{\alpha_3}+\psi_{s,d}^{\text{inf}} (\mathfrak{C}_k)^{\alpha_3}\right]^{\frac{1}{\alpha_3}}.
\end{split}
\end{align*}
Thus, ~\eqref{eqn:eqlabel27dd} holds.\\
{\textbf{Lemma 2.2.7.}} Suppose that $s>d$, and $\mathfrak{C} \subset \mathbb{R}^d$ is a compact set with $0<\mu(\mathfrak{C})<\infty$, $\mathcal{T}(d) > 2$ is continuous and derivative for  $d \in \mathbb{R}^+$. Furthermore, suppose that for any compact subset $\mathfrak{C}_{i} \subset \mathfrak{C}$, the limit $\psi_{s,d}(\mathfrak{C}_{i}), i \in \mathbb{R}^+$ exists and is given by 
\begin{equation*}\label{eqn:eqlabelaa2a}
\psi_{s,d}(\mathfrak{C}_{i})=\frac{{\color{ red}C_8}}{\mathcal{U}_d^s(\mathfrak{C}_{i})^{\mathcal{T}(d)-2}}.
\end{equation*}
Then, $\psi_{s,d}(\mathfrak{C})$ exists and is given by
\begin{equation}\label{eqn:eqlabelaa2a2}
\psi_{s,d}(\mathfrak{C})=\frac{{\color{ red}C_8}}{\mathcal{U}_d^s(\mathfrak{C})^{\mathcal{T}(d)-2}}.
\end{equation}
Moreover, if a sequence of $n$-point configurations $X_{1:n}$ is asymptotically weighted Riesz energy minimizing on the set $\mathfrak{C}$ and $\mu(\mathfrak{C})>0$, then
\begin{equation}\label{eqn:eqlabel1a4}
\lim_{n\rightarrow \infty }\frac{1}{n}\sum_{i=1,i\neq j}^{n}\parallel x_i-x_j\parallel \rightarrow u_d^s(\mathbb{S}).
\end{equation}
{\textbf{Proof}} To prove ~\eqref{eqn:eqlabelaa2a2}, we firstly decompose the entire metric space $\mathfrak{C}$ into extremely small disconnected parts with diameter less than $\epsilon>0$, according to the property of Borel metrics, then
\begin{equation}\label{eqn:eqlabel1a4sa}
\sum_{P\in \mathfrak{C}_{i}}\mathcal{U}_d(P)\leq \mathcal{U}_d(\mathfrak{C}).
\end{equation}
Hereafter we follow an argument in \cite{borodachov2019discrete} and define a sufficiently small space $\mathfrak{C}_{i}$ as follows, the rule refers to \cite{borodachov2008asymptotics}.
We consider the hyperplane $\mathfrak{C}'$ consisting of all points, $(-l,l)$ is a cube embedded in $\mathfrak{C}'$, we discretize the cube with tiny intervals for $j$-th ordinate, $-l= h_0^j<h_1^j\cdots <h_k^j= l,j\in i^{d'},d'\in(1,d),i=(i_1,i_2\cdots,i_n)$, $k$ is sufficiently large, $\exists \left \| h_k^j-h_{k-1}^j \right \|< \epsilon$ such that ~\eqref{eqn:eqlabel1a4sa} holds. $\mathfrak{C}_{i}$ can be written as
\begin{equation*}\label{eqn:eqlabel1a4asa}
\mathfrak{C}_{i}:=[ h_{i_1^1}^1,h_{i_1^1+1}^1)\times \cdots \times[h_{i_n^{d'}-1}^{d'},h_{i_n^{d'}}^{d'}),
\end{equation*}
For $\mathfrak{C}_{i}\subset \mathfrak{C}$, if $\omega (x_i,x_j)$ is bounded, let 
\begin{equation*}\label{eqn:eqlabel1a4as7a}
\overline{\omega}_{{\mathfrak{C}}_{i}}=\sup_{x_i,x_j \in \mathfrak{C}_{i}}\omega(x_i,x_j), \text{ and} \ \underline{\omega}_{{\mathfrak{C}}_{i}}=\inf_{x_i,x_j \in \mathfrak{C}_{i}}\omega(x_i,x_j),
\end{equation*}
we introduce the radial basis functions $\varphi(\cdot)$ to approximate the corresponding bounded $\omega(x_i,x_j)$:
\begin{equation}\label{eqn:eqlabel1a4asa2a}
\overline{\omega}_{{\mathfrak{C}}_{i}}(x_i,x_j)=\sum_{P\in \mathfrak{C}_{i}}\overline{\omega}_P\varphi (\left \| x_i-x_j \right \|),\ \ \underline{\omega}_{{\mathfrak{C}}_{i}}(x_i,x_j)=\sum_{P\in \mathfrak{C}_{i}}\underline{\omega}_P\varphi (\left \| x_i-x_j \right \|).
\end{equation}
From Lemma 2.2.5, and ~\eqref{eqn:eqlabel27},
\begin{align*}
\label{eqn:eqlabel2aa}
\begin{split}
\psi_{s,d}^{{\text{sup}}} (\mathfrak{C})^{\alpha3} &\geq \sum_{i=1}^n\psi_{s,d}^{{\text{sup}}} (\mathfrak{C}_i)^{\alpha3}\geq \sum_{i=1}^n\left [ \overline{\omega}_{{\mathfrak{C}}_{i}}(x_i,x_j)\cdot \psi_{s,d}^{{\text{sup}}} (\mathfrak{C}_i) \right ]^{\alpha3}={{\color{ red}C_8}}\sum_{x_i,x_j\in \mathfrak{C}_{i}}\overline{\omega}_{{\mathfrak{C}}_{i}}^{\alpha3}\cdot \mathcal{U}_d(\mathfrak{C}_i)\\
&\geq {{\color{ red}C_8}}\int_{x_i,x_j \in \mathfrak{C}_{i}}\overline{\omega}_{{\mathfrak{C}}_{i}}(x_i,x_j)^{\alpha3}d\mathcal{U}_d(\mathfrak{C}_i).
\end{split}
\end{align*}
From Lemma 2.2.6 and ~\eqref{eqn:eqlabel27dd}, similarly, we have
\begin{align*}
\begin{split}
\psi_{s,d}^{{\text{inf}}} (\mathfrak{C})^{\alpha3} &\leq \sum_{i=1}^n\psi_{s,d}^{{\text{inf}}} (\mathfrak{C}_i)^{\alpha3}\leq \sum_{i=1}^n\left [ \underline{\omega}_{{\mathfrak{C}}_{i}}(x_i,x_j)\cdot \psi_{s,d}^{{\text{inf}}} (\mathfrak{C}_i) \right ]^{\alpha3}={{\color{ red}C_8}}\sum_{x_i,x_j\in \mathfrak{C}_{i}}\underline{\omega}_{{\mathfrak{C}}_{i}}^{\alpha3}\cdot \mathcal{U}_d(\mathfrak{C}_i)\\
&\leq {{\color{ red}C_8}}\int_{x_i,x_j \in \mathfrak{C}_{i}}\underline{\omega}_{{\mathfrak{C}}_{i}}(x_i,x_j)^{\alpha3}d\mathcal{U}_d(\mathfrak{C}_i).
\end{split}
\end{align*}
Given a sufficiently small $P$, for ~\eqref{eqn:eqlabel1a4asa2a}, use the equation limit, we have
\begin{align*}
\begin{split}
& \overline{\omega}_{{\mathfrak{C}}_{i}}(x_i,x_j)=\sum_{P\in \mathfrak{C}_{i}}\overline{\omega}_P\varphi (\left \| x_i-x_j \right \|)={\omega}_{{\mathfrak{C}}}(x_i,x_j),\\
&\underline{\omega}_{{\mathfrak{C}}_{i}}(x_i,x_j)=\sum_{P\in \mathfrak{C}_{i}}\underline{\omega}_P\varphi (\left \| x_i-x_j \right \|)={\omega}_{{\mathfrak{C}}}(x_i,x_j).
\end{split}
\end{align*}
Since $\omega(x_i,x_j)$ is continuous on $\mathfrak{C}$, both $\int_{x_i,x_j \in \mathfrak{C}_{i}}\overline{\omega}_{{\mathfrak{C}}_{i}}(x_i,x_j)^{\alpha3}d\mathcal{U}_d(\mathfrak{C}_i)$ and $\int_{x_i,x_j \in \mathfrak{C}_{i}}\underline{\omega}_{{\mathfrak{C}}_{i}}(x_i,x_j)^{\alpha3}d\mathcal{U}_d(\mathfrak{C}_i)$ converge to $\mathcal{U}_d^s(\mathfrak{C})$. Consequently, 
 the limit $\psi_{s,d}(\mathfrak{C}_{i}), i \in \mathbb{R}^+$ exists and can be given by 
\begin{equation*}
\psi_{s,d}(\mathfrak{C}_{i})=\frac{{\color{ red}C_8}}{\mathcal{U}_d^s(\mathfrak{C}_{i})^{\mathcal{T}(d)-2}}.
\end{equation*}
By the Fatou's Lemma and Monotone Convergence Theorem, 
Thus,  ~\eqref{eqn:eqlabelaa2a2} holds on $\mathfrak{C}$.

To prove ~\eqref{eqn:eqlabel1a4}, suppose that $X_{1:n}$ is an asymptotically weighted Riesz energy minimizing sequence of $n$-point configuration on $\mathfrak{C}$, the corresponding signed finite Borel measures $\cup_{i=1}^n{\mu_d^s(\mathfrak{C}_i)}$ in $\mathbb{R}^d$ converges weak$^*$ to a signed finite Borel measure $\mu_d(\mathfrak{C})$, as $n \rightarrow \infty$. Consequently, ~\eqref{eqn:eqlabel1a4} is equivalent to the assertion that 
\begin{equation*}\label{eqn:eqlabel1a4a2aa}
\lim_{n\rightarrow \infty}\sum_{j=1}^{n}p_j=\cup_{i=1}^n \mu_d^s(\mathfrak{C}_j)=\mu_d(\mathbb{S})
\end{equation*}
holds for any almost $\sigma\text{-algebra}$ subset on $\mathfrak{C}$, let $\mathfrak{C}_{\sigma}= \cup_{i=1}^n \mathfrak{C}_i$ be a subset of $\sigma\text{-algebra}$ on $\mathfrak{C}$,
for any Borel subset $\mathfrak{C}_{\sigma}\subset \mathfrak{C}$. Since $\mathfrak{C}_{\sigma}$ and $\mathfrak{C}/\mathfrak{C}_{\sigma}$ are the compact subsets of $\mathfrak{C}$, suppose
$\psi_{s,d}(\mathfrak{C}_{\sigma})=\frac{{\color{ red}C_9}}{\mu(\mathfrak{C}_{\sigma})^{-\frac{1}{\alpha_3}}}$ and $\psi_{s,d}(\mathfrak{C}/\mathfrak{C}_{\sigma})=\frac{{\color{ red}C_{10}}}{\mu(\mathfrak{C}/\mathfrak{C}_{\sigma})^{-\frac{1}{\alpha_3}}}$, for the asymptotically weighted Riesz energy minimal sequence $X_{1:n}$, 
\begin{equation*}
\label{eqn:eqlabel10huj}
\begin{split}
 \lim_{n\rightarrow \infty }\frac{E^s (X_{1:n})}{\lambda (n)}& ={\color{red} C_{s,d}} \cdot (\mu(\mathfrak{C}))^{\frac{1}{\alpha_3}} = {\color{red} C_{s,d}} \cdot (\mu(\mathfrak{C}_{\sigma})+\mu(\mathfrak{C}/\mathfrak{C}_{\sigma}))^{\frac{1}{\alpha_3}} \\
  & = \left [ \psi_{s,d}(\mathfrak{C}_{\sigma})^{\alpha_3}+\psi_{s,d}(\mathfrak{C}/\mathfrak{C}_{\sigma})^{\alpha_3})\right ]^\frac{1}{\alpha_3}.
\end{split}
\end{equation*}
Using ~\eqref{eqn:eqlabel2adfjn} in Lemma 2.2.6 and ~\eqref{eqn:eqlabelaa2a2} which holds for $\mathfrak{C}_{\sigma}$ and $\mathfrak{C}/\mathfrak{C}_{\sigma}$, we have
\begin{equation*}
\label{eqn:eqlabel10x32}
\begin{split}
 \lim_{n\rightarrow \infty}\sum_{j=1}^{n}p_j& =\frac{\psi_{s,d}(\mathfrak{C}/\mathfrak{C}_{\sigma})^{-\alpha_3}}{\psi_{s,d}(\mathfrak{C}_{\sigma})^{-\alpha_3}+\psi_{s,d}(\mathfrak{C}/\mathfrak{C}_{\sigma})^{-\alpha_3}} =\frac{\mathcal{U}_d^s(\mathfrak{C}_{\sigma})}{\mathcal{U}_d^s(\mathfrak{C}_{\sigma})+\mathcal{U}_d^s(\mathfrak{C}/\mathfrak{C}_{\sigma})}=\mu_d(\mathbb{S}).
\end{split}
\end{equation*}

Thus, ~\eqref{eqn:eqlabel1a4} holds.
\subsection{Separation Distance}
In this section, we focus on the features of points generated by minimizing the weighted Riesz energy criterion from the point of discrete geometry. We state and prove the relationship between asymptotically best-packing and weighted Riesz energy-minimizing configuration first. We are devoted to deriving the lower bounds for the minimal pairwise separation in optimal configurations. Suppose that $\mathfrak{C}$ is a compact infinite space with Euclidean distance $r$, $\mathfrak{C} \times \mathfrak{C} \rightarrow [0,\infty )$, and define the separation distance of an $n$-point configuration $X_{1:n}=\{x_1,x_2,...,x_n\}$ on $(r,\mathfrak{C})$ as $r_{\text{min}}(X_{1:n}):=\min_{1\leq i\neq j\leq n}\parallel x_i-x_j\parallel,$
the $n$-point best-packing configuration on $\mathfrak{C}$ is $X_{1:n}^*$ satisfying
\begin{align}
\label{eqn:eqlabel10x}
\begin{split}
  & \lim_{m\rightarrow \infty }{(\cup X_{1:n}^{(s)})}=\mathfrak{C}, r_{\text{min}}^{\text{max}}(\mathfrak{C}):=\max\{r_{\text{min}}(X_{1:n }^*),X_{1:n}^* \subset \mathfrak{C}\}.
\\
\end{split}
\end{align}
For the weighted Riesz energy criterion, the $n$-point best-packing configuration is the limiting case of $s$ approximating infinity. 
To derive the well-separateness property, we will first provide the upper estimate of the Borel measure in a restricted compact space for the potential and then prove the estimate for our weighted Riesz energy configurations next.\\
$\textbf{Lemma 2.3.1.}$ Let $\mu$ be a closed Borel measure in $\mathbb{R}^d$, $\epsilon >0$, suppose $B(y,\epsilon) 
\subset \sigma\text{-algebra}$, there exist $\mu(\sigma\text{-algebra}) \leq \epsilon$, $y\in \mathfrak{C}_i^\infty$, then the potential
\begin{equation}
\label{eqn:eqlabel10x3}
\begin{split}
  G_{s,\mu}(y,\epsilon)& :=\left[ \int _{\mathfrak{C} \setminus B(y,\epsilon)}\frac{\omega(x_i,y)}{\parallel x_i-y\parallel^s} d\mu(x_i) \right]^\frac{1}{s} \leq \sum_{k'=1}^{\infty}{{\color{ red}C_{11}}}{\epsilon}^{\frac{2d}{s(2d+k'!)}-1}.
\end{split}
\end{equation}

\textbf{Proof of Lemma 2.3.1} For every $y\in \mathfrak{C}$ and $\epsilon>0$, the point $x_i=y$ is enclosed in an infinite dimensional space (denoted by sphere) of radius $\epsilon$, in order to make the energy function  (7) in the main text 
integrable, this space will be excluded with $\mathfrak{C} \setminus B(y,\epsilon)$. Substitute ~\eqref{eqn:eqlabel20ew} into (7) in the main text, 
 then
\begin{align}
\label{eqn:eqlabel10x4}
\begin{split}
  G_{s,\mu}^s(y,\epsilon) :&=\int _{\mathfrak{C} \setminus B(y,\epsilon)}h_{\beta}(x_i,y) d\mu(x_i) =\int _0^{{\epsilon}^{-s}}\mu\{x\in \mathfrak{C} \setminus B(y,\epsilon):h_{\beta_1^{+}}(x_i,y)>t\}dt \\
  & = \int _0^{{\epsilon}^{-s}}\mu\{x\in \mathfrak{C} \setminus B(y,\epsilon):\frac{e^{\left [ \beta_1 \parallel x_i-y\parallel \right ]^{-\frac{s}{2d}}}}{\parallel x_i-y\parallel^s}>t\}dt \\
  & =  \sum_{k'=1}^{\infty}\int _0^{{\epsilon}^{-s}}\mu\{x\in \mathfrak{C} \setminus B(y,\epsilon):\frac{\left[\beta_1 \cdot \parallel x_i-y\parallel\right]^{\frac{-{s{k'}}}{2d}}}{{k'}!\parallel x_i-y\parallel^s}>t\}dt \\
  & \leq \sum_{k'=1}^{\infty}\int _0^{{\epsilon}^{-s}}\mu\left[B(y,\frac{k'}{{\beta_1}^{\frac{-sk'}{2d}}}t^{\frac{-2d}{2sd+sk'!}})\right]dt\\
  &\leq \sum_{k'=1}^{\infty}\frac{k'}{{\beta_1}^{\frac{-sk'}{2d}}}\int _0^{{\epsilon}^{-s}}t^{\frac{-2d}{2sd+sk'!}}dt\\
  &\leq \sum_{k'=1}^{\infty}\frac{k'}{{\beta_1}^{\frac{-sk'}{2d}}}\frac{2sd+sk'!}{2sd+sk'!-2d}{\epsilon}^{\frac{2d}{2d+k'!}-s}\\
  & \leq \sum_{k'=1}^{\infty}{{\color{ red}C_{10}}}{\epsilon}^{\frac{2d}{2d+k'!}-s}.
\end{split}
\end{align}
Thus, Lemma 2.3.1 holds.

Let $v$ the county measure of the intersection of compact subsets $\mathfrak{C}_i$ and $\mathfrak{C}_j$, there exists an infinitely small value $\zeta > 0$ such that $v=\lim_{i,j\in [1,\infty )}\#(\mathfrak{C}_i\cap \mathfrak{C}_j)> \zeta$, and $\lim_{i,j\in [1,\infty )}\left \| x_i-x_j \right \|\subset \mu (\mathfrak{C}_i^\infty \cap \mathfrak{C}_j^\infty)\subset  \sigma\text{-algebra}$. Suppose that $\overline{W}$ be the bounded value of $\omega(x_i,x_j)$ in ~\eqref{eqn:eqlabel10}, we can estimate the upper bound for extremal weighted Riesz energy criterion.\\
\textbf{Lemma 2.3.2.} Let $ s\geq d> 0$, there exist constant ${\color{ red}C_{12}}$ and ${\color{ red}C_{13}}$ such that for any compact set $\mathfrak{C} \subset \mathbb{R}^d$ with the measure $\mu(\sigma\text{-algebra})>0$ and any $\omega(x_i,x_j)$ that is bounded and lower semicontinuous on $\mathfrak{C} \times \mathfrak{C}$, then 
\begin{equation*}\label{eqn:eqlabel44}
\mathcal{E}_{\beta} (\mathfrak{C} ,n) \leq \frac{{\color{ red}C_{12}}\overline{W}}{\mu(\sigma\text{-algebra})}n^{\frac{1}{d}+\frac{2}{s}}. 
\end{equation*}
holds for any weighted Riesz energy minimizing configuration $X_{1:n}^*, s\geq 2$, and 
\begin{equation}\label{eqn:eqlabel442}
r_{\text{min}}(X_{1:n}^*)\geq {\color{ red}C_{13}} \cdot\mu(\sigma\text{-algebra})^{\frac{2d}{2d+1}} \cdot n^{\frac{-1}{d}+\frac{-2}{s}}. 
\end{equation}
\textbf{Proof of Lemma 2.3.2} Let $E_i(x)=\left \{\sum_{i=1}^{n-1}\sum_{j=i+1}^{n}\frac{\omega(x,x_j)}{\parallel x-x_j\parallel^s} \right \}^{\frac{1}{s}},x \in \mathfrak{C}$. If we define $x_i$ is from the unit of minimum energy configuration, $E_i(x_i)\leq E_i(x),x\in \mathfrak{C}, i=1,...,n$. If $\mu$ is a measure on $\mathfrak{C}$ where $x_j$ is excluded by $\mathfrak{C}_{\setminus{x_j}}:=\mathfrak{C} \setminus \sum_{j\neq i}(\cup B(x_j,\epsilon))$, as $\epsilon$ is extremely small, there exists a constant  ${\color{ red}C_{14}}$ that is close to 1, $0<{\color{ red}C_{14}}<1$, $\mu(\mathfrak{C}_{\setminus{x_j}})\geq \mu(\mathfrak{C})-\mu(\sum_{j\neq i}(\cup B(x_j,\epsilon)))\geq {\color{ red}C_{14}} \cdot \mu(\mathfrak{C})$. When $s>d$, substitute (7) in the main text 
into $E_i^s(x_i)$, we have
\begin{align}
\label{eqn:42d}
\begin{split}
  E_i^s(x_i)  &\leq \frac{2}{\mu(\mathfrak{C}_{\setminus{x_j}})}\int _{\mathfrak{C}_{\setminus{x_j}}}\int _{\mathfrak{C}_{\setminus{x_j}}}h_{\beta}(x_i,x_j)d\mu(x_i) d\mu(x_j) \\
  &= \frac{2}{\mu(\mathfrak{C})}\int _{\mathfrak{C}_{\setminus{x_j}}}\int _{\mathfrak{C}_{\setminus{x_j}}}\frac{\omega(x,x_j)}{\parallel x-x_j\parallel^s} d\mu(x) d\mu(x_j)= \frac{2}{\mu(\mathfrak{C}_{\setminus{x_j}})}\sum_{j:j\neq i}G_{s,\mu}^s(x_j,\epsilon)\\
  & \leq \frac{1}{{\color{ red}C_{15}} \cdot \mu(\mathfrak{C}_{\setminus{x_j}})} \sum_{j:j\neq i}\sum_{k'=1}^{\infty}\frac{k'}{{\beta_1}^{\frac{-sk'}{2d}}}\frac{2sd+sk'!}{2sd+sk'!-2d}{\epsilon}^{\frac{2d}{2d+k'!}-s}\\
  &\leq {{\color{ red}C_{15}}}\left(\frac{n}{\mu(\mathfrak{C})} \right)^{\frac{s}{d}+2}\leq \left[\frac{{\color{ red}C_{16}}\overline{W}}{\mu(\sigma\text{-algebra})}n\right]^{\frac{s}{d}+2}.
\end{split}
\end{align}
Let $X_{1:n}^*,n\geq 2$, be a weighted Riesz energy minimizing configuration on $\mathfrak{C}$ and $i_n$ and $j_n$ be such that $r_{\text{min}}(X_{1:n}^*)=\left \| x_i^*-x_j^* \right \| $ and for the bound $\left \| \omega(x_i,x_j) \right \| \leq \overline{W}$, then
\begin{equation}
\label{eqn:eqlabel44q}
\mathcal{E}_{\beta} (\mathfrak{C} ,n) \geq \sum_{j=1,j\neq i}^{n}\left[h_{\beta}(x_i,x_j)\right ]^{\frac{1}{s}}\geq  \frac{1}{\left \| x_i^*-x_j^* \right \|}.
\end{equation}

Combine ~\eqref{eqn:42d} and ~\eqref{eqn:eqlabel44q}, we have
\begin{equation*}\label{eqn:eqlabel424q}
\frac{1}{\left \| x_i^*-x_j^* \right \|}\leq \mathcal{E}_{\beta_1} (\mathfrak{C} ,n)\leq \left[\frac{{\color{ red}C_{12}}\overline{W}}{\mu(\sigma\text{-algebra})}n\right]^{\frac{1}{d}+\frac{2}{s}},
\end{equation*}
then, $r_{\text{min}}(X_{1:n}^*)\geq {\color{ red}C_{13}} \cdot\mu(\sigma\text{-algebra})^{\frac{2d}{2d+1}} \cdot n^{\frac{-1}{d}+\frac{-2}{s}}$.
Consequently, Lemma 2.3.2 holds.

\subsection{\color{ red}Covering Radius for journal}


In this section, we state and prove the bound of the covering radius. And extend to deal with the weak* limit distribution of best-covering $n$-point configurations on rectifiable sets $\mathfrak{C}$. Suppose that $\mathfrak{C}$ is bounded in a compact infinite metric space with Euclidean metric $r$, $\mathfrak{C} \times \mathfrak{C} \rightarrow [0,\infty )$, we generalize $ r_{\text{min}}^{\text{max}}(\mathfrak{C})$ in ~\eqref{eqn:eqlabel10x} and define the covering radius of an $n$-point configuration $X_{1:n}$ in a metric space $(\mathfrak{C},r)$ as $\rho(X_{1:n},\mathfrak{C}):= \max_{x \in \mathfrak{C}}\min_{i=1,...,n}r(x,x_i).$
From the geometrical perspective, the covering radius of $X_{1:n}$ can be considered as the minimal radius of $n$ adjacent closed balls centered at $X_{1:n}$ whose union contains the entire $\mathfrak{C}$. Among finite element analysis and approximation theory, this quantity is known as the best approximation of the set $\mathfrak{C}$ by the configuration $X_{1:n}$ \cite{borodachov2019discrete}. The optimal values of this quantity are also of interest and we define the minimal $N$-point covering radius of a set $\mathfrak{C}$ as
\begin{equation*}\label{46b}
\rho_n(\mathfrak{C}):=\min\{\rho(X_{1:n},\mathfrak{C}): X_{1:n} \subset \mathfrak{C}\}.
\end{equation*}
$\rho_n(\mathfrak{C})$ is also called an $n$-point best-covering configuration for $\mathfrak{C}$ \cite{hardin2012quasi}.\\
 \textbf{Theorem 2.4.1.} Assume that $s>d$ and $\mathfrak{C} \subset \mathbb{R}^d$, if the compact support of $\mu$ is contained in $\mathfrak{C}$, with respect to some finite Borel measure for a compact Euclidean metrical and bounded set, there exists positive constant ${\color{ red}C_{17}}$ such that every sequence $X_{1:n}^*$ of weighted Riesz energy minimizing configurations on $\mathfrak{C}$ satisfies
\begin{equation}\label{eqn:eqlabel44b}
 \rho{(X_{1:n}^*)} \leq {\color{ red}C_{17}}  n^{(1-\frac{s}{d})\cdot \frac{2d+1}{s+2d+2ds} }
\end{equation}
\textbf{Proof of Theorem 2.4.1} Let $X_{1:n}^*=\{x_1^*...,x_n^*\}$ be an $n$-point energy minimizing configuration for the compact and measurable space $\mathfrak{C}$, we consider the function
\begin{equation*}\label{eqn:eqlabel191}
\overline{H}(y):=\frac{1}{n}\sum_{i=1}^n\frac{ \omega(x_i,y)}{\parallel x_i-y\parallel^s}.
\end{equation*}
Given a specified $1 \leq j \leq n$ and $d>0$ in $\mathfrak{C}$, 
\begin{equation}\label{eqn:eqlabel192}
\overline{H}(y):=\frac{1}{n}\frac{ \omega(x_j,y)}{\parallel x_j-y\parallel^s}+\frac{1}{n}\sum_{i=1,i \neq j}^n\frac{ \omega(x_i,y)}{\parallel x_i-y\parallel^s}.
\end{equation}
since $X_{1:n}^*$ is the energy minimizing configuration on $\mathfrak{C}$, the point $x_j$ minimizes ~\eqref{eqn:eqlabel192}, consequently, for each fixed $j$ and $y\in \mathfrak{C}$,
\begin{equation}\label{eqn:eqlabel1966}
\overline{H}(y)\geq \frac{1}{n}\frac{ \omega(x_j,y)}{\parallel x_j-y\parallel^s}+\frac{1}{N}\sum_{i=1,i\neq j}^n \frac{ \omega(x_j,y)}{\parallel x_j-y\parallel^s}.
\end{equation}
Summing over ~\eqref{eqn:eqlabel1966} with different $j$, from Jensen's inequality, it gives
\begin{align*}
\label{eqn:eqlabel194}
\begin{split}
  n \cdot \overline{H}(y) & \geq H(y)+\frac{1}{n}\sum_{i=1}^n\sum_{j=1,j\neq i}^n \frac{\omega(x_j,x_i)}{\parallel x_i-x_j\parallel}\geq \overline{H}(y)+\frac{1}{n}E^s(x_i,x_j),
\end{split}
\end{align*}
thus,
\begin{equation}\label{eqn:eqlabel1793}
\overline{H}(y)\geq \frac{1}{n(n-1)}E^s(x_i,x_j)\geq \frac{E^s(x_i,x_j)}{n^2}, (y \in \mathfrak{C}).
\end{equation}
Since $\mathfrak{C}$ is compact, there exists a point $y^* \in \mathfrak{C}$ such that
\begin{equation}\label{eqn:eqlabel19io3}
\min_{x \leq i \leq n}r(y^*,x_i)=\rho(X^*_{1:n},\mathfrak{C}).
\end{equation}
 $\omega$ is bounded on $\mathfrak{C}$, by Theorem 2.2.1, there are constants $N_0$ and ${\color{ red}C_{18}}>0$ such that
 \begin{equation*}\label{eqn:eqlabeld19ido3}
E^s(x_i,x_j)\geq \mathcal{E}_{\beta} \left [(\mathfrak{C} ,n) \right]^s={\color{ red}C_{18}}n^{2+\frac{s}{d}}, n\geq N_0.
\end{equation*}
 Since ~\eqref{eqn:eqlabel1793} holds for the point $y^*$ of ~\eqref{eqn:eqlabel19io3}, substitute ~\eqref{eqn:eqlabel19io3} into ~\eqref{eqn:eqlabel1793}, 
 \begin{equation}\label{eqn:eqlabel179d3}
\overline{H}(y)\geq \frac{\mathcal{E}_{\beta} \left [(\mathfrak{C} ,n) \right]^s}{n^2}={\color{ red}C_{18}}n^{\frac{s}{d}}, (y \in \mathfrak{C}).
\end{equation}
 Next we determine an upper bound for $\overline{H}(y)$  on $\mathfrak{C}$ by introducing ~\eqref{eqn:eqlabel19io3}. Lemma 2.3.3 applies that there exist some ${\color{ red}C_{19}}>0$ such that $ r_{\text{min}}(X_{1:n}^*) \geq {\color{ red}C_{19}} \cdot\mu(\sigma\text{-algebra})^{\frac{2d}{2d+1}} \cdot n^{\frac{-1}{d}+\frac{-2}{s}}$ for $n\geq 2$.  
Thus, we have
\begin{equation}\label{eqn:eqlabel179dd3}
 \rho(X_{1:n}^*,\mathfrak{C}) \geq {\color{ red}C_{20}} \cdot\mu(\sigma\text{-algebra})^{\frac{2d}{2d+1}} \cdot n^{\frac{-1}{d}+\frac{-2}{s}}
\end{equation}
Hereafter we follow an argument in \cite{borodachov2019discrete} and define
 \begin{equation*}\label{eqn:eqlabeld7}
 \rho_0:= \zeta \cdot   \rho(X_{1:n}^*,\mathfrak{C}) 
\end{equation*}
 where $\zeta$ is sufficiently small, the particles are well-separated that have no intersection with the neighbor within the Ball $B(x_i,r_0) \subset \mathfrak{C}$.
 For any $x\in B(x,x_i)$, combine ~\eqref{eqn:eqlabel19io3} and ~\eqref{eqn:eqlabel179dd3}
\begin{align}
\label{eqn:eqlabeld71}
\begin{split}
   r(x,y^*) & \leq r(x,x_i)+r(x_i,y^*) \leq \rho_0+r(x_i,y^*) \leq \zeta \rho(X_{1:N}^*)+r(x_i,y^*) \leq (1+\zeta)r(x_i,y^*),
\end{split}
\end{align}
the lower bound for $r(x,y^*)$
\begin{align}
\label{eqn:eqlabeld72}
\begin{split}
   r(x,y^*) & \geq r(x,x_i)-r(x,x_i) \geq r(x_i,y^*)-\rho_0 \geq  r(x_i,y^*)-\zeta \rho(X_{1:n}^*) \geq (1-\zeta) \rho(X_{1:n}^*),
\end{split}
\end{align}
implies
 \begin{equation}\label{eqn:eqlabeldp7}
{\bigcup_{i=1}^{n}}B(x_i,\rho_0)\subseteq \mathfrak{C} (y^*,(1-\zeta)\rho(X_{1:n}^*)).
\end{equation}
For fixed $n$ points, using ~\eqref{eqn:eqlabeld71} and taking the average value on $B(x_i,\rho_0)$, we obtain
\begin{align*}
\begin{split}
   \overline{K}(y)& \leq \frac{\parallel \omega \parallel(1+\zeta)^s{\color{ red}C_{21}}}{n\rho_0} \sum_{i=1}^{n}\int _{B(y,\epsilon)}\frac{ d\mu(x_i) }{\parallel x-y^*\parallel^s}\\
   & \leq \frac{(1+\zeta)^s{\color{ red}C_{21}}}{n\rho_0}\int _{\mathfrak{C} \setminus B(y^*,(1-\zeta)\rho(X_{1:n}^*))}\frac{ \omega(x,y^*)d\mu(x_i) }{\parallel x-y^*\parallel^s}\\
   & = \frac{(1+\zeta)^s{\color{ red}C_{21}}}{n\rho_0}\int _{\mathfrak{C} \setminus B'}h_{\beta}(x,y^*)d\mu(x_i),
\end{split}
\end{align*}
where $B'=B(y^*,(1-\zeta)\rho(X_{1:n}^*))$, let $c(k')={\frac{-2d}{2sd+sk'!}}$, we decompose the right-hand side of ~\eqref{eqn:eqlabeldp7} into two terms and proceed as in ~\eqref{eqn:eqlabel10x4} to obtain
\begin{align}
\label{eqn:eqlabeld75b}
\begin{split}
   \int _{\mathfrak{C} \setminus B'}h_{\beta}(x,y^*)d\mu(x_i)
   & = \int _{B'}h_{\beta}(x,y^*)d\mu(x_i)+\int _{\mathfrak{C} \setminus B(y^*,\rho_0)}h_{\beta}(x,y^*)d\mu(x_i)\\
  &\leq {\color{ red}C_{22}}\int _0^{\rho_0^{-s}}t^{c(k')}dt+\sum_{k'=1}^{\infty}\frac{k'}{{\beta_1}^{\frac{-sk'}{2d}}}\int _{\rho_0^{-s}}^{{[(1-\zeta)\rho(X_{1:n}^*)]}^{-s}}t^{c(k')}dt\\
  &= {\color{ red}C_{22}}\cdot c'(k')+\sum_{k'=1}^{\infty}\frac{k'}{{\beta_1}^{\frac{-sk'}{2d}}}\left[ \frac{[(1-\zeta){\rho{(X_{1:n}^*)}}]^{c(k')s-s}}{c(k')s-s} -c(\rho_0)\right]\\
  & \leq {\color{ red}C_{23}} n \cdot [(1-\zeta){\rho{(X_{1:n}^*)}}]^{\frac{-2ds-2d-s}{2d+1}},
  \end{split}
\end{align}
where $c'(k')= \frac{\rho_0^{c(k')s-s}}{c(k')s-s}$.
Combine ~\eqref{eqn:eqlabel179d3} with ~\eqref{eqn:eqlabeld75b}
\begin{equation*}\label{eqn:eqlabel179detg5}
 \frac{\mathcal{E}_{\beta} \left [(\mathfrak{C} ,n) \right]^s}{n^2}={\color{ red}C_{18}}n^{\frac{s}{d}}, (y \in \mathfrak{C}) \leq \overline{H}(y)\leq {\color{ red}C_{23}} n \cdot [(1-\zeta){\rho{(X_{1:n}^*)}}]^{\frac{-2ds-2d-s}{2d+1}} .
\end{equation*}
Thus, we have
\begin{equation*}\label{eqn:eqlabel179de5}
 \rho{(X_{1:n}^*)} \leq {\color{ red}C_{17}}  n^{(\frac{s}{d}-1)\cdot \frac{2d+1}{-s-2d-2ds} }
\end{equation*}
~\eqref{eqn:eqlabel44b} {\color{ red}in the main text} holds.
\section{Weighted Riesz Particles MCMC}

In this section, we develop a new sampler in which the propagation of particles originates from weighted Riesz energy minimization, called weighted Riesz particles, which inherits the special properties introduced in Section 2 when traversing through discrete deterministic subdomains of the parameter space via particle interactions. We further extend this to sequential sampling in the particle Metropolis-Hastings framework for the inference of hidden Markov models, where the acceptance rate is approximated by the pseudo-marginal Metropolis-Hastings algorithm.

\subsection{Sequential Weighted Riesz Particles Sampling}

Finding the optimal design of a configuration is non-deterministic, especially for high dimensions, where point-by-point traversal leads to exponential growth of the computational load. Many optimization algorithms have been proposed for the optimal design of different configurations. Park \cite{park1994optimal} proposed a two-stage exchange and Newton-type optimal design that minimizes the integrated mean square error and maximizes the entropy, respectively. Ye~\cite{ye1998orthogonal} further extended the column-pair algorithm. Morris and Mitchell~\cite{morris1995exploratory} adapted simulated annealing \cite{kirkpatrick1983optimization} to explore cells in the reachable domain. 
Inspired by \cite{morris1995exploratory} and \cite{joseph2015sequential}, we propose a constrained one-point-at-a-time greedy algorithm for developing sequential designs of weighted Riesz particles as follows.

(\uppercase\expandafter{\romannumeral1}) The selection of the initial point is critical because it is closely related to the sampling of subsequent points. For the sake of
stability of the initial point, we use the particle with the largest mean value as the initial point. Our desired value is 
$\mathbb{E}
(x)=\int_0^{x}xf(x)dx, x \in \mathfrak{C}$. The maximum point  $x_0$ can be obtained by $x_0=\text{arg}_x[\max \mathbb{E}
(x)]$.\\
(\uppercase\expandafter{\romannumeral2}) Given an initial point $x_0$, we can sequentially generate $x_2,x_3...,x_n$. Suppose we have $n$ points using ~\eqref{eqn:eqlabel10}. Then the $(n+1)$th point can be obtained by
\begin{align}
\label{eqn:eqlabel10q}
\begin{split}
 x_{n+1} & =\underset{x}{\text{arg}} \mathcal{E}_{\beta}(\mathfrak{C},n) =\text{arg}\min_x \left \{ \sum_{i=1}^{n-1}\sum_{j=i+1}^{n}\frac{\omega(x_i,x_j)}{\parallel x_i-x_j\parallel^s}  \right \}^{\frac{1}{s}}. 
 \end{split}
\end{align}
(\uppercase\expandafter{\romannumeral3}) If $\left | x_{n+1}-x_n \right |\ge r_{\text{min}}(x_{1:n}^*)$, we further develop an acceptance criterion for $x_{n+1}$: Given $u\sim U(u\mid 0,1)$, if $\frac{\left | x_{n+1}-x_n \right |}{\left |x_n \right |}\geq u$, we accept $x_{n+1}$; otherwise, we reject it.\\
(\uppercase\expandafter{\romannumeral4}) From $n$ points, we can use some statistical methods such as regression or kriging to estimate the latent manifold, where the density can be updated with the ancestors of the samples, and then recursively generated for different configurations of discrete manifolds.

\subsection{Pseudo-marginal Metropolis-Hastings Sampling}
Consider a hidden Markov model, where $x_t \sim f_{\theta}(x_t \mid x_{t-1}),y_t\mid x_t \sim g_{\theta}(y_t \mid x_t),$ given $\hat{x_0}$, $x_t (t=1,2,...n)$ is a latent variable to be observed, and assuming that the measurements ${y_t}$ are conditionally independent of the given ${x_t}$, the objective is to estimate $\{x_{1:t}, \theta\}$. The Particle Metropolis-Hastings \cite{andrieu2010particle}, proposed an MCMC method to randomly explore in the assumed measurable $\theta$ space and generate samples from the approximated posterior $\hat{p}(x_{1:t},\theta \mid y_{1:t})$, whose closed-form ${p}(x_{1:t},\theta \mid y_{1:t}) = {p}(\theta \mid y_{1:t}) \cdot {p}(x_{1:t} \mid y_{1:t}, \theta) $ is not reachable for exact pointwise evaluation. 

We will introduce how weighted Riesz particles are used for the following steps: In the parameter space, given $\{\theta,x_{1:t}\}$, a new status $\{\theta',x_{1:t}^{'}\}$ is obtained from a proposal $q(\theta',x_{1:t}^{'}\mid \theta,x_{1:t})$ with the probability of acceptance
\begin{align}
\label{eqn:eqlabel1011q}
\begin{split}
\alpha & =\min\left \{ 1,\frac{p(x_{1:t}^{'},\theta'\mid y_{1:t})q(\theta,x_{1:t}\mid \theta',x_{1:t}^{'})}{p(x_{1:t},\theta\mid y_{1:t})q(\theta',x_{1:t}^{'}\mid\theta,x_{1:t} )} \right \}= \min\left \{ 1,\frac{p(y_{1:t}\mid \theta')p(\theta^{'})q(\theta\mid \theta')}{p(y_{1:t}\mid \theta)p(\theta)q(\theta'\mid \theta)} \right \}.
 \end{split}
\end{align}
The optimal importance density function that minimizes the variance of importance weights, conditioned upon $x_{t-1}^i$ and ${y_t}$ has been shown \cite{doucet2000sequential} to be 
\begin{align*}
\label{eqn:eqlabel1011aq}
\begin{split}
q(x_t\mid x_{t-1}^i,y_t)_{opt} & =p(x_t\mid x_{t-1}^i,y_t)  =\frac{p(y_t\mid x_t,x_{t-1}^i)p(x_t\mid x_{t-1}^i)}{p(y_t\mid x_{t-1}^i)}.
 \end{split}
\end{align*}
While sampling from $p(y_t\mid x_t,x_{t-1}^i)$ may not be straightforward, since $x_{1:t}$ belongs to the "deterministic" part of the discrete manifolds of the space, $x_{1:t} \in \mathfrak{C}$, 
the selection of importance density $ q(x_t|y_t, x_{t-1}^{a_{t-1}^{(i)}})$ is from the configuration of the minimum energy, where $a_{t}^{(i)}$ denotes the ancestor of particle $x_t^i$
. If $n\rightarrow \infty$, we have $\lim_{n\rightarrow \infty}q(x_t|y_t,x_{t-1}^{a_{t-1}^{(i)}})=p(x_t|x_{t-1}^i,y_t)$.
Thus, our proposal converges to the optimal importance density. We can obtain a stochastic estimator of $p(y_{1:T}\mid \theta)$ and this likelihood can be estimated by the weights 
\begin{equation}\label{eqn:eqlabel119s3a}
\hat{p}_\theta(y_{1:T})=\prod_{t=1}^T(\frac{1}{N_x}\sum_{i=1}^{N_x}\frac{p(x_t|x_{t-1}^i,y_t)}{q(x_t|y_t,x_{t-1}^{a_{t-1}^{(i)}})}).
\end{equation}
It can be shown that $\mathbb{E}[\hat{p}_\theta(y_{1:T})]=p_{\theta}(y_{1:T})$ \cite{chopin2004central}. The variance of the weights will approximate to $0$ when $N_x$ goes to infinity, this would be verified by the following experiments.

From ~\eqref{eqn:eqlabel1011q} and ~\eqref{eqn:eqlabel119s3a}, we can get the estimated acceptance ratio
\begin{equation*}\label{eqn:eqlaabel1319s3a}
\hat{\alpha}= \min\left \{ 1,\frac{\hat{p}(y_{1:t}\mid \theta')p(\theta^{'})q(\theta\mid \theta')}{\hat{p}(y_{1:t}\mid \theta)p(\theta)q(\theta'\mid \theta)} \right \}.
\end{equation*}
\section{Experiments}

In this section, we present simulations embedding Riesz particles into sequential Monte Carlo, as well as extending it to Bayesian analysis of linear and nonlinear models. We ran the experiments on an HP Z200 workstation with an Intel Core i5 and an $\#82-18.04.1-$ Ubuntu SMP kernel. The code is available at \url{https://github.com/986876245/Weighted-Riesz-Particles}.

\subsection{Linear Gaussian State Space Model}


The expression for the linear model is
\begin{equation*}\label{linearModel}
X_t\mid X_{t-1} \sim g(X_t|X_{t-1}),\ \ \
Y_t\mid X_t \sim f(Y_t|X_{t})+e_o.
\end{equation*}
Where $g(X_t|X_{t-1})=\phi X_{t-1}+e_v$, the noise from tracking $e_v \sim \mathbb{N}(0,\delta _v^2)$, the noise from observations $e_o \sim \mathbb{N}(0,\delta_o^2)$. Here we use ~\eqref{eqn:eqlabel10q}, to compute $ \widehat{x}^n_{0:T}$, and $\widehat{p}^n_\theta(y_{1:T})$ with Riesz particles by
\begin{align*}
\label{eqn:eqlabel10g}
\begin{split}
 g(\widehat{x}_t|\widehat{x}_{t-1}) & =\text{arg}\min_x \mathcal{E}_{\beta}(\mathfrak{C},N') =\text{arg}\min_x \left \{ \sum_{x\neq \widehat{x}_i,i=1}^{t-1}\frac{\omega(\widehat{x}_i,x)}{\parallel \widehat{x}_i-x\parallel^s}  \right \}^{\frac{1}{s}},\\
  \omega (\widehat{x}_i,x)&\propto e^{\left [ \kappa (\widehat{x}_i) \kappa (x) +\beta \parallel \widehat{x}_i-x\parallel \right ]^{-\frac{s}{2d}}}.
\end{split}
\end{align*}
For a linear Gaussian state-space model, the optimal proposed distribution of propagating particles $X^i_t, i=1,...n$ is derived \cite{dahlin2015getting} from
\begin{equation*}
\label{increasess}
\begin{split}
       p_{\theta}^{\text{opt}}(X_t^i \mid X_{t-1}^i,Y_t) & \propto  g_{\theta}(Y_t \mid Y_t^i)f_{\theta}(X_t^i \mid X_{t-1}^i) =\mathbb{N}(X^i_t;\sigma^2[\sigma_o^{-2}Y_t+\sigma^{-2}_v\phi X^i_{t-1}],\sigma^2)
\end{split}    
 \end{equation*}
 with $\sigma^{-2}=\sigma_v^{-2}+\sigma_o^{-2}$. To ensure the stability of the algorithm and to minimize the variance of the incremental particle weights for the current time step, we set $\kappa (x_t) \propto p_{\theta}^{\text{opt}}(X_t^i \mid X_{t-1}^i,Y_t)$. The latent state $x_t$ can be represented with an unbiased quantity $\widehat{x}^n_t=\frac{1}{n}\sum_{i=1}^{n}x^i_t$, here, $n$ is the number of particle to estimate the current state, $N'$ is the quantity of Riesz particles to discretize the submanifolds. Ensuring a uniform embedding of the particles with labels, we allocate the indices one by one with the remainder of $n$ divided by $N',(n>N')$ for the specific particles $x^i_t$.

\begin{figure*}[!htp]
	\centering
		\centering
		\includegraphics[width=.95\linewidth]{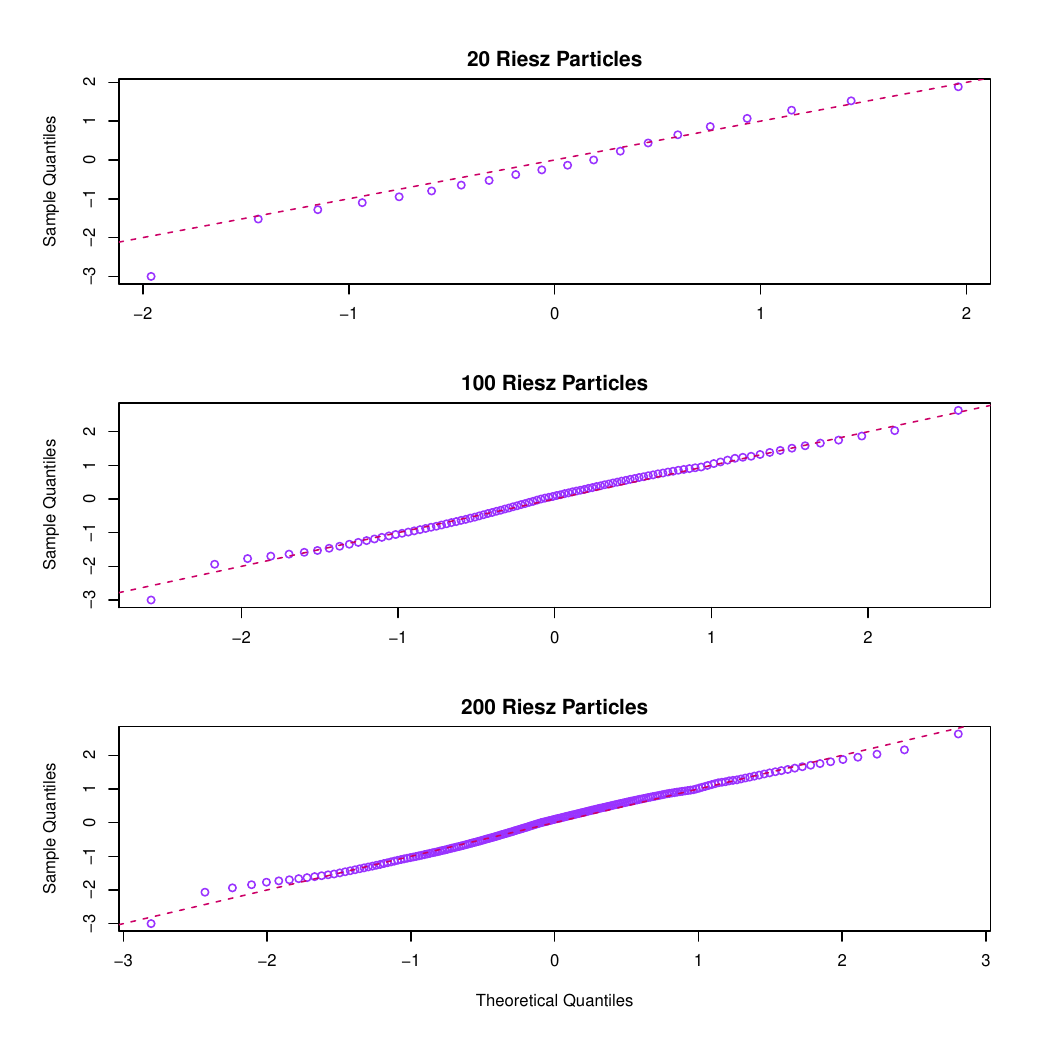}
  		\caption{Theoretical Quantiles for 20, 100 and 200 Riesz particles.}
		\label{fig:experiment2a4g}
	
\end{figure*}

We first study the generation of sets of particles with different cardinality in the same space under the same conditions to test and verify the geometric characteristics. When these particles are mapped to a particular low-dimensional space shown in \autoref{fig:experiment2a4g}, they will approach different straight lines with the target parameters of the Ritz particles as $\{\beta=2,m=40,d=1\}$, it satisfies the uniform distribution, which is verified Theorem 2.2.1.
We then embed these particles in a sequential Monte Carlo, where these states interact recursively in the set of Riesz particles set $\mathbb{P}$, and for comparison with the ground truth, we provide a simulated data record of the model, with observations of $T=250$, initial values of ${\phi=0.75,\delta_v=1.00,\delta_o= 0.10}$, $\widehat{x}_0=0$. We recorded the estimated logarithmic bias and logarithmic MSE for Riesz particles embedded in sequential Monte Carlo at different particle sets with different cardinality, respectively. It is shown in ~\autoref{tab:table_particles}.

\begin{figure*}[!htp]
	\centering
		\centering
        \includegraphics[width=.95\linewidth]{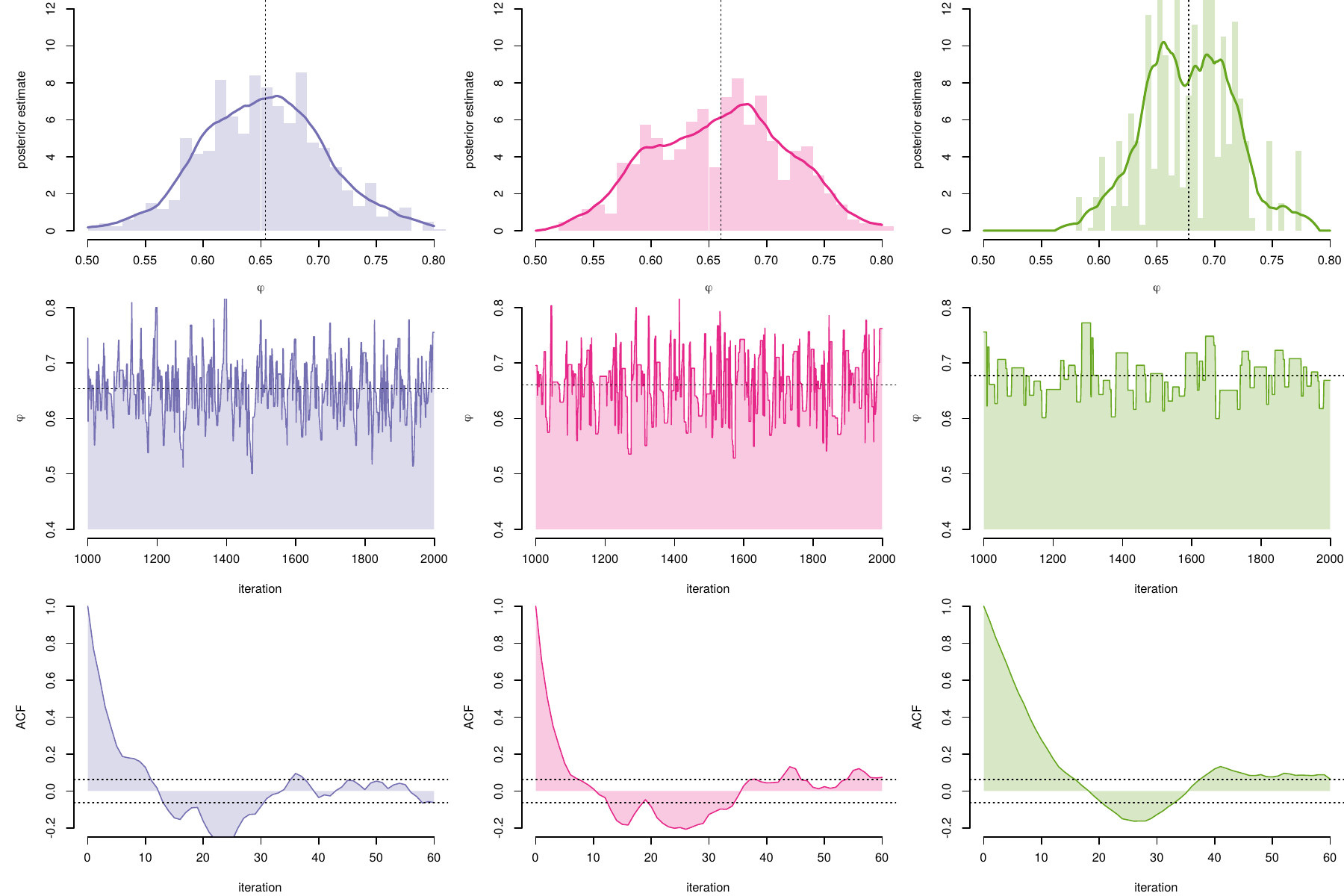}
		\caption{Posterior estimate, burning process and ACF for different step size:$h_1=0.05, h_2=0.1, h_3=0.5$.}
		\label{fig:experiment334}
	
\end{figure*}
\begin{table*}
\begin{center}
\begin{tabular}{|c|c|c|c|c|c|c|c|}
\hline
Number of particles(n) & 10 & 20 & 50 & 100 & 200 & 500 & 1000 \\
\hline\hline
log-bias & -3.67 & -4.05 & -4.46 & -4.87 & -5.24 & -5.64 & -5.97 \\
log-MSE & -6.94 & -7.64 & -8.43 & -9.27 & -9.98 & -10.81 & -11.44 \\
\hline
\end{tabular}
\end{center}
\caption{The log-bias and the log-MSE of the filte red states under 100 Riesz particles for varying n.}
\label{tab:table_particles}
\end{table*}
Here, we extend the weighted Riesz particles to the pseudo-marginal Metropolis-Hastings algorithm for Bayesian parameter inference of hidden Markov models provided in Section 3.2. We estimate $\phi$, $\phi \in (-1,1)$ describing the persistence of the state and keeping ${delta_v=1.00, \delta_e=0.10}$ fixed, and $\phi_0=0.75$ a prior, and specify the number of Riesz particles in the set $\mathbb{P}$ to be $100$, which is much smaller than the number of iterations ($\geq 2000$), and from this point on we have essentially scaled the particle set of the model, which then requires very little evaluation for us to infer. We performed iterations with different step sizes, $h_1=0.05,h_2=0.1,h_3=0.5$, and the lagged autocorrelation plots of the posterior estimates, the combustion process, are shown in \autoref{fig:experiment334}. The corresponding table is shown in \autoref{tab:table_particles2}.

\begin{table*}
\begin{center}
\begin{tabular}{|c|c|c|c|c|c|c|}
\hline
Number of observations(T) & 10 & 20 & 50 & 100 & 200 & 500 \\
\hline\hline
Estimated posterior mean & 0.587 & 0.775 & 0.745 & 0.712 & 0.689 & 0.722 \\
Estimated posterior variance & 0.031 & 0.016 & 0.011 & 0.009 & 0.005 & 0.002 \\
\hline
\end{tabular}
\end{center}
\caption{The estimated posterior mean and variance when varying T.}
\label{tab:table_particles2}
\end{table*}

\subsection{Nonlinear State Space Model}
We continue by presenting a practical application of our proposal for tracking stochastic volatility, a nonlinear state-space model with Gaussian noise, in which log volatility is considered as a latent variable and an important factor in financial risk management analysis. The stochastic volatility is given by
\begin{align}
\label{eqn:eqlabel10gasa}
\begin{split}
&x_0\sim \mathbb{N}(\mu,\frac{\sigma _v^2}{1-\rho  ^2}),\ \ x_t\mid x_{t-1} \sim  \mathbb{N}(\mu+\rho(x_{t-1}-\mu),\sigma _v^2),\ \ y_t\mid x_t \sim  \mathbb{N}(0,exp(x_t)\tau),
\end{split}
\end{align}
where the parameters $\theta =\left \{\mu, \rho  ,\sigma_v,\tau \right \}$, $\mu\in \mathbb{R},
\rho  \in [-1,1]$, $\sigma _v $ and $\tau \in \mathbb{R}_+$, denotes the mean value,
persistence of volatility, standard deviation of the state process, and instantaneous volatility, respectively.

\begin{figure*}[!htp]
	\centering
		\centering
 		\includegraphics[width=.95\linewidth]{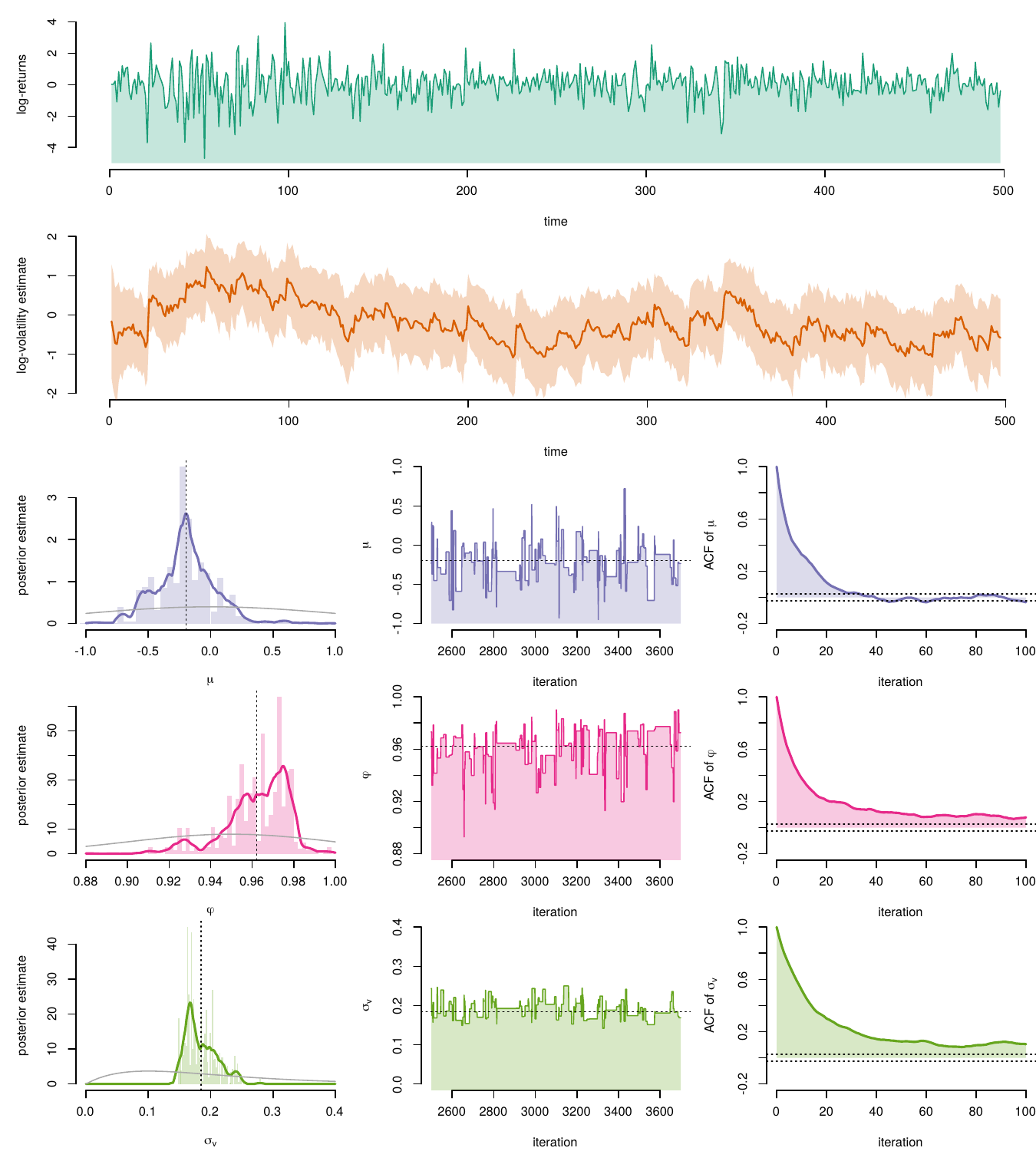}
		\caption{Top: The daily log-returns and estimated log-volatility with $95\%$ confidence intervals of the NASDAQ OMXS30 index for the period between February 4, 2015 and February 4, 2016. Bottom: the posterior estimate(left), the trace of the Markov chain(middle) and the corresponding ACF(right) of $\mu$(purple), $\phi$(magenta) and $\sigma_v$(green) obtained from Riesz particles embedded PMH. The dotted and solid gray lines in the left and middle plots indicate the parameter posterior mean and the parameter priors, respectively.}
		\label{fig:experimentu4aa}
	
\end{figure*}

The observation $y_t=\log(p_t/p_{t-1})$,  also called log-returns, represents the logarithm of the daily difference in the exchange rate $p_t$, 
here, $\{{p_t}\}_{t=1}^T$ is the daily closing prices of the NASDAQ OMXS30 index (a weighted average of the 30 most traded stocks on the Stockholm stock exchange) \cite{dahlin2015getting}. We extract the data from {\color{blue} \href{https://www.quandl.com/}{Quandl}} for the period between January 2, 2015 and January 2, 2016. The resulting logarithmic returns are shown in \autoref{fig:experimentu4aa}. Large fluctuations are frequent, which is known as volatility clustering in finance, and from equation ~\eqref{eqn:eqlabel10gasa}, volatility clustering effect is more likely to occur when $|\phi|$ is close to $1$ and the standardized variance is very small. Here, the parameters of the objective for Riesz particles: $\{\beta=2,m=40,d=1\}$, the size of Riesz particles is $180$. The initial values are $\mu_0=0, \sigma_0=1, \phi_0=0.95, \sigma_{\phi}=0.05, \delta_{v_0}=0.2, \sigma_{v}=0.03$ 
We obtain good performance, inferring a posteriori estimates after only a few evaluations, greatly reducing the computational load of high-dimensional sampling,
as shown in \autoref{fig:experimentu4aa}.

\section{Conclusion}
Markov Chain Monte Carlo (MCMC) provides a viable method for the inference of hidden Markov models, but it tends to be computationally overloaded, especially by the curse of dimensionality, as the Monte Carlo sampler traverses the uncertain region of the parameter space in small random steps. In the process, a large number of duplicate samples are burned, and these duplicate samples increase the computational load significantly. We introduce a deterministic sampling mechanism where all generated samples are produced by particle interactions under a weighted Riesz energy minimization criterion. All samples inherit the properties of good separation and bounded coverage radius. We embed it into MCMC, obtain high performance through experiments with Hidden Markov Models, require only a small number of evaluations, and we can extend the method to high-dimensional sampling.
For future research, we will propose some kernel for the Riesz particles and scale the model with low complexity of computations from the perspective of equilibrium states on high-dimensional sampling.

\section*{Acknowledgments}
This was supported in part by BRBytes project.

\bibliographystyle{unsrt}  
\bibliography{references}

\begin{thebibliography}{10}

\bibitem{hastings1970monte}
W~Keith Hastings.
\newblock Monte carlo sampling methods using markov chains and their applications.
\newblock 1970.

\bibitem{metropolis1953equation}
Nicholas Metropolis, Arianna~W Rosenbluth, Marshall~N Rosenbluth, Augusta~H Teller, and Edward Teller.
\newblock Equation of state calculations by fast computing machines.
\newblock {\em The journal of chemical physics}, 21(6):1087--1092, 1953.

\bibitem{geman1984stochastic}
Stuart Geman and Donald Geman.
\newblock Stochastic relaxation, gibbs distributions, and the bayesian restoration of images.
\newblock {\em IEEE Transactions on pattern analysis and machine intelligence}, (6):721--741, 1984.

\bibitem{gelfand1990sampling}
Alan~E Gelfand and Adrian~FM Smith.
\newblock Sampling-based approaches to calculating marginal densities.
\newblock {\em Journal of the American statistical association}, 85(410):398--409, 1990.

\bibitem{robert1999monte}
Christian~P Robert, George Casella, and George Casella.
\newblock {\em Monte Carlo statistical methods}, volume~2.
\newblock Springer, 1999.

\bibitem{gelman1997weak}
Andrew Gelman, Walter~R Gilks, and Gareth~O Roberts.
\newblock Weak convergence and optimal scaling of random walk metropolis algorithms.
\newblock {\em The annals of applied probability}, 7(1):110--120, 1997.

\bibitem{breger2018points}
Anna Breger, Martin Ehler, and Manuel Gr{\"a}f.
\newblock Points on manifolds with asymptotically optimal covering radius.
\newblock {\em Journal of Complexity}, 48:1--14, 2018.

\bibitem{brandolini2010quadrature}
Luca Brandolini, Christine Choirat, Leonardo Colzani, Giacomo Gigante, Raffaello Seri, and Giancarlo Travaglini.
\newblock Quadrature rules and distribution of points on manifolds.
\newblock {\em arXiv preprint arXiv:1012.5409}, 2010.

\bibitem{damelin2005point}
Steven~B Damelin and V~Maymeskul.
\newblock On point energies, separation radius and mesh norm for s-extremal configurations on compact sets in rn.
\newblock {\em Journal of Complexity}, 21(6):845--863, 2005.

\bibitem{hardin2012quasi}
Douglas~P Hardin, Edward~B Saff, and J~Tyler Whitehouse.
\newblock Quasi-uniformity of minimal weighted energy points on compact metric spaces.
\newblock {\em Journal of Complexity}, 28(2):177--191, 2012.

\bibitem{hardin2005minimal}
DP~Hardin and EB~Saff.
\newblock Minimal riesz energy point configurations for rectifiable d-dimensional manifolds.
\newblock {\em Advances in Mathematics}, 193(1):174--204, 2005.

\bibitem{borodachov2008asymptotics}
S~Borodachov, D~Hardin, and E~Saff.
\newblock Asymptotics for discrete weighted minimal riesz energy problems on rectifiable sets.
\newblock {\em Transactions of the American Mathematical Society}, 360(3):1559--1580, 2008.

\bibitem{borodachov2014low}
Sergiy~V Borodachov, Douglas~P Hardin, and Edward~B Saff.
\newblock Low complexity methods for discretizing manifolds via riesz energy minimization.
\newblock {\em Foundations of Computational Mathematics}, 14(6):1173--1208, 2014.

\bibitem{fekete1923verteilung}
Michael Fekete.
\newblock {\"U}ber die verteilung der wurzeln bei gewissen algebraischen gleichungen mit ganzzahligen koeffizienten.
\newblock {\em Mathematische Zeitschrift}, 17(1):228--249, 1923.

\bibitem{smale1998mathematical}
Steve Smale.
\newblock Mathematical problems for the next century.
\newblock {\em The mathematical intelligencer}, 20(2):7--15, 1998.

\bibitem{hardin2004discretizing}
DP~Hardin, EB~Saff, et~al.
\newblock Discretizing manifolds via minimum energy points.
\newblock {\em Notices of the AMS}, 51(10):1186--1194, 2004.

\bibitem{hunter2011measure}
John~K Hunter.
\newblock Measure theory.
\newblock {\em University Lecture Notes, Department of Mathematics, University of California at Davis. http://www. math. ucdavis. edu/\~{} hunter/measure\_theory}, 2011.

\bibitem{borodachov2019discrete}
Sergiy~V Borodachov, Douglas~P Hardin, and Edward~B Saff.
\newblock {\em Discrete energy on rectifiable sets}.
\newblock Springer, 2019.

\bibitem{park1994optimal}
Jeong-Soo Park.
\newblock Optimal latin-hypercube designs for computer experiments.
\newblock {\em Journal of statistical planning and inference}, 39(1):95--111, 1994.

\bibitem{ye1998orthogonal}
Kenny~Q Ye.
\newblock Orthogonal column latin hypercubes and their application in computer experiments.
\newblock {\em Journal of the American Statistical Association}, 93(444):1430--1439, 1998.

\bibitem{morris1995exploratory}
Max~D Morris and Toby~J Mitchell.
\newblock Exploratory designs for computational experiments.
\newblock {\em Journal of statistical planning and inference}, 43(3):381--402, 1995.

\bibitem{kirkpatrick1983optimization}
Scott Kirkpatrick, C~Daniel Gelatt, and Mario~P Vecchi.
\newblock Optimization by simulated annealing.
\newblock {\em science}, 220(4598):671--680, 1983.

\bibitem{joseph2015sequential}
V~Roshan Joseph, Tirthankar Dasgupta, Rui Tuo, and CF~Jeff Wu.
\newblock Sequential exploration of complex surfaces using minimum energy designs.
\newblock {\em Technometrics}, 57(1):64--74, 2015.

\bibitem{andrieu2010particle}
Christophe Andrieu, Arnaud Doucet, and Roman Holenstein.
\newblock Particle markov chain monte carlo methods.
\newblock {\em Journal of the Royal Statistical Society: Series B (Statistical Methodology)}, 72(3):269--342, 2010.

\bibitem{doucet2000sequential}
Arnaud Doucet, Simon Godsill, and Christophe Andrieu.
\newblock On sequential monte carlo sampling methods for bayesian filtering.
\newblock {\em Statistics and computing}, 10(3):197--208, 2000.

\bibitem{chopin2004central}
Nicolas Chopin.
\newblock Central limit theorem for sequential monte carlo methods and its application to bayesian inference.
\newblock {\em The Annals of Statistics}, 32(6):2385--2411, 2004.

\bibitem{dahlin2015getting}
Johan Dahlin and Thomas~B Sch{\"o}n.
\newblock Getting started with particle {M}etropolis-{H}astings for inference in nonlinear dynamical models.
\newblock {\em Journal of Statistical Software}, 88(2):1--41, 2019.

\end{thebibliography}

\end{document}